\documentclass[debug]{rmaa}

\usepackage{paralist}

\usepackage{psfrag,color}

\usepackage[latin1]{inputenc}

\usepackage{amsmath}
\usepackage{graphics}
\usepackage{soul}
\usepackage{xcolor}

\newcommand{\hchii}{\hbox{HC{ }H{ }{II}{ }}}
\newcommand{\uchii}{\hbox{UC{ }H{ }{II}{ }}}
\newcommand{\hii}{\hbox{H{ }{II}{ }}}
\newcommand{\hiir}{\hbox{H{ }{II}{ }}region}
\newcommand{\hiirs}{\hbox{H{ }{II}{ }}regions}
\newcommand{\uchiir}{UC~\hii~region}

\newcommand\asec   {\ifmmode{^{\prime \prime}}\else{$^{\prime \prime}$}\fi}
\newcommand\pasec  {\ifmmode{{\rlap.}^{\prime \prime}}\else{${\rlap.}^{\prime \prime}$}\fi}
\newcommand\gax    {\ifmmode{_>\atop^{\sim}}\else{${_>\atop^{\sim}}$}\fi}

\title{Search and study of ultracompact \hii regions}

\author{
  D. Quiroga-Gonz\'alez\altaffilmark{1,2},
  M. A. Trinidad\altaffilmark{1,3}, 
  E. de la Fuente\altaffilmark{4},
  J. M. Masqu\'e\altaffilmark{1}, and
  T. Rodr\'iguez--Esnard\altaffilmark{5}
  }

\altaffiltext{1}{Departamento de Astronom\'ia, Universidad de Guanajuato, Guanajuato, M\'exico.}

\altaffiltext{2}{Universidad Aut\'onoma Metropolitana, M\'exico.}

\altaffiltext{3}{Corresponding author}

\altaffiltext{4}{Departamento de F\'isica, CUCEI, Universidad de Guadalajara, Jalisco, M\'exico.}

\altaffiltext{5}{Instituto de Geof\'{i}sica y Astronom\'{i}a de la Republica de Cuba}

\shortauthor{Quiroga-Gonz\'alez, Trinidad, de la Fuente, et al. }
\shorttitle{Search and study of ultracompact \hii regions}

\fulladdresses{
\item Daniel Quiroga-Gonz\'alez: Universidad Aut\'onoma Metropolitana-Iztapalapa, M\'exico
(dquirogag@gmail.com).

\item Miguel Angel Trinidad: Departamento de Astronom\'ia, Universidad de Guanajuato, Apdo. Postal 144, 36000 Guanajuato, M\'exico 
(trinidad@ugto.mx).

\item Eduardo de la Fuente: Departamento de F\'isica, CUCEI, Blvd. Gral. Marcelino Garc\'{i}a Barrag\'an 1421, Ol\'{i}mpica, 44430, Guadalajara, Jalisco, M\'exico 
(eduardo.delafuente@academicos.udg.mx).

\item Josep M. Masqu\'e: Departamento de Astronom\'{i}a, Universidad de Guanajuato, Apartado Postal 144, 36000, Guanajuato, Guanajuato, M\'exico
(jmasque@ugto.mx).
  
\item Tatiana Rodr\'iguez--Esnard: Instituto de Geof\'{i}sica y Astronom\'{i}a de la Republica de Cuba, calle 212 \#~2906 e/ 29 y 31 la Coronela, la Lisa 11600, C. Habana, Cuba (tatiana@iga.cu) 
(esnard73@gmail.com).

}

\listofauthors{W. J. Henney, A. Collaborator, \& L. Author}
\indexauthor{Henney, W. J.}
\indexauthor{Collaborator, A.}
\indexauthor{Author, L.}

\abstract{We present results from a sample of 106 high-luminosity IRAS sources observed with the Very Large Array in the B and C configurations.  96 sources were observed in the X-band and 52 in the K-band, with 42 of them observed at both wavelengths. We also used previously published observations in the C-band for 14 of them. The detection rate of sources with 3.6~cm  continuum emission was $\sim25\%$, while only 10\% have emission at 1.3~cm. In order to investigate the nature of these sources, their physical parameters were calculated mainly using the 3.6~cm continuum emission, and for sources detected at two wavelengths, we used the best fit of three \hii region models with different geometries. As a final result, we present a catalog of the detected sources, which includes their basic physical parameters for further analysis. The catalog contains 17  ultracompact \hii regions and 3 compact \hii regions.}

\resumen{Presentamos resultados de una muestra de 106 fuentes IRAS de alta luminosidad observadas con el interfer\'ometro Very Large Array en las configuraciones B y C. 96 fuentes se observaron en la banda X y 52 en la banda K, con 42 de ellas observadas en ambas longitudes de onda. Tambi\'en usamos observaciones previamente publicadas en la banda C para 14 de ellas. La tasa de detecci\'on de fuentes con emisi\'on de continuo a 3.6~cm fue del $\sim25\%$,  mientras que s\'olo un 10\% tienen  emisi\'on a 1.3~cm. Para investigar la naturaleza de estas fuentes, se calcularon sus par\'ametros f\'isicos usando principalmente la emisi\'on de continuo a 3.6~cm y para fuentes detectadas en dos longitudes de onda usamos el mejor ajuste de tres modelos de regiones \hii  con diferentes geometr\'ias. Como resultado final, presentamos un cat\'alogo de las fuentes detectadas, proporcionando sus par\'ametros f\'isicos b\'asicos para su posterior an\'alisis. El cat\'alogo contiene 17  regiones \hii ultracompactas y 3 regiones \hii compactas.}

\addkeyword{ISM: \hii regions}
\addkeyword{ISM: General}
\addkeyword{Stars: formation}
\addkeyword{Radio continuum: ISM}
\addkeyword{Instruments: VLA}
\addkeyword{Metods: Models}

\begin{document}
\maketitle

\section{Introduction}
\label{sec:intro}

The study of high-mass star formation is crucial for understanding the physical and chemical evolution of galaxies. Because forming massive stars takes $\sim 10^6$~yr, the process of high mass star formation is less understood than the formation of low-mass stars (time-scales of $\sim 10^9$~yr). Currently, it is not completely clear how massive stars form, being monolithic collapse, protostar collision and/or coalescence, and competitive accretion the most widely accepted models \citep[see][and references therein]{Motte-Bontemps-Louvet(2018)}. Studying the evolution of the earliest phases of high-mass star formation is key to understanding how this process occurs. In this sense, two of the earliest phases of massive star formation are the young stellar object and the \hiir \citep[e.g][]{Garay1999}. While both stages have been extensively studied,  new questions continue to arise about its formation and evolution process. Further study of these evolutionary phases will undoubtedly contribute to a better understanding of massive star formation and provide evidence for or against the proposed models.

One way to contribute to solving the puzzle of high-mass star formation is to study the \hii regions related to this process: the hyper-compact (HC), ultra-compact (UC), and compact \hii regions.  These objects are thought to be related to the evolutionary sequence as the massive star approaches the zero-age main sequence or ZAMS \citep[e.g][and references therein]{Beuther2007}. The Physical parameters that define \hchii, \uchii, and compact HII regions following this evolutionary sequence are shown in Table \ref{tab:tab1}. 

On the other hand, density gradients are highly noticeable in \hii regions \citep[e.g.][]{DePree1995,Jaffe1999,Franco2000a,Franco2000b,Franco2001,Phillips2007,Phillips2008}. These gradients are important and useful to describe the dynamics of a \hii region. For example, density gradients with a power law of n$_{\rm e}$ $\propto$ r$^{\beta}$, where r is the distance from the ionization front, accurately describe expanding \hii regions when $\beta \gtrsim$ 1.5  \citep[e.g][and references therein]{Franco1990,Franco2000a,Franco2000b,Franco2001}. Thus, the presence of these gradients should be taken into consideration in models and studies of \hchii, \uchii, and compact \hii regions.

In order to advance in the understanding of the earliest stages of the high-mass star formation process and to find evidence in favor of one of the models mentioned above, we perform a physical characterization of the ionized gas in a sample of 106 IRAS sources to identify \hiirs  { }in their different evolutionary stages. We calculate physical parameters at 3.6 cm in the standard way, and we apply density gradient models for sources with multiple wavelength observations. We aim to confirm if they are \hiirs, and if applicable, to determine their nature and classify them as \hchii, \uchii, compact \hiir, taking into consideration the presence of protostellar thermal jets.

The sample, radio continuum observations, and data reduction are described in \S~\ref{sec:sec2}. Results and discussion are presented in \S~\ref{sec:sec3} and \S~\ref{sec:sec4}, respectively. Finally, we give the conclusions in \S~\ref{sec:sec5} and individual sources comments are provided in Appendix \ref{sec:ap-A}.

\section{Observations}
\label{sec:sec2}
We retrieve 3.6 and 1.3~cm data for a sample of 104 IRAS sources from the Very Large Array (VLA\footnote{The National Radio Astronomy Observatory is a facility of the National Science Foundation operated under cooperative agreement by Associated Universities, Inc.}) archive using the B and C configurations, respectively (unpublished data from the AC295 project; P.I. Ed Churchwell).  Out of the 104 sources, 94 were observed in the X band and 52 in the K band, with 42 observed at both wavelengths. We also included two sources (IRAS 18094-1823 and G45.47+0.05) observed in the X band with the VLA D configuration (AK559 project; P.I. Stan Kurtz; see \citet{delaFuente2018,delaFuente2020a}), bringing the final sample to 106 sources. The observations of the AC295 project, at both wavelengths, were carried out in snapshot mode using a bandwidth of 50~MHz, with an integration time of 5 and 10 minutes for 3.6 and 1.3~cm, respectively, over a time span of about 4.5 months (1992 January and May) for both bands. Table \ref{tab:tab2} lists the 106 sources in the sample observed at 3.6 and 1.3~cm. Additionally, we also used 6~cm data, observed with the VLA-B and reported by \citet{Urquhart2009}, for some sources detected at 3.6 and/or 1.3~cm. All sources in the sample are located in star-forming regions associated with high luminosity IRAS sources (L$_{\rm FIR}$ $\gtrsim 500$~L$_{\odot}$), and are situated more than 1~kpc away. They cover a range from $\sim 21^h$ to 08.5$^h$ in right ascension (J2000), and from $\sim -41^{\circ}$ to 66$^{\circ}$ in declination (J2000).  These characteristics make the sources in the sample excellent candidates for identifying and studying HII regions, as well as for expanding the dataset of these objects to better understand their properties. Additional information about each source can be found in Appendix \ref{sec:ap-A}.

We performed the data editing, calibration, and further mapping of all sample sources at 3.6 and 1.3~cm wavelengths following the standard techniques using the Common Astronomy Software Applications (CASA) of the NRAO version 5.3.0-143 \citep{McMullin2007}. The flux calibrator for observations at 3.6 and 1.3~cm was 3C48, and several phase calibrators were used (see Table \ref{tab:tab_2}).  6~cm data were calibrated using the same procedure as was used for the 3.6 and 1.3~cm data. In order to obtain a similar angular resolution for continuum sources detected at two and three wavelengths, we convolved the data with the same beam. All observational parameters of the detected sources (position, flux density, and deconvolved angular size) were obtained with the task IMFIT of CASA.

From the subsample of  96 IRAS sources observed at 3.6~cm, we detect only 25 of them, while from the subsample of 52 sources observed at 1.3~cm, we detect only five. These five sources were also detected at 3.6~cm (see Table \ref{tab:tab3}). The low detection rate may be due to the low sensitivity of the observations carried out in snapshot mode, but other reasons cannot be rule out (see subsection \ref{sec:sec4.2}).

Observational parameters of all detected IRAS sources at 1.3, 3.6, and 6~cm are listed in Table \ref{tab:tab4}, and their respective radio contour maps are shown in Figures \ref{fig:fig1}, \ref{fig:fig2},  \ref{fig:fig3}, and \ref{fig:fig4}. Detailed results for each of the sources are provided in Appendix \ref{sec:ap-A}.

\section{Results}
\label{sec:sec3}

\subsection{3.6~cm Continuum Emission: Physical parameters}
\label{sec:sec3.1}
In order to investigate the nature of the radio continuum sources detected toward the IRAS regions, we used the 3.6~cm flux density to determine their physical parameters as if they were optically thin \hii regions at this wavelength. We also assume homogeneous and isothermal gas, with an spherically symmetric distribution, composed of pure hydrogen and  a canonical value for the electronic temperature of $10^4$~K. The electronic density (n$_{\rm e}$), emission measure (EM), the mass of the ionized gas (M$_{\rm HII}$), and the total rate of Lyman continuum photons of the ionizing star (N$'_{\rm c}$) were calculated in the standard way using equations from \ref{eqn:eqn1} to \ref{eqn:eqn4} \citep[][]{Schraml1969,Kurtz1994}:

\begin{equation}
\begin{split}
\label{eqn:eqn1}
\left(\frac{\mathrm{n_e}}{\mathrm{cm^{-3}}}\right) = 7.8\times 10^3 \left( \frac{\nu}{\mathrm{4.9\, GHz}}\right)^{0.05}
\left(\mathrm{\frac{S_\nu}{mJy}}\right)^{0.5}
\left(\frac{\mathrm{T_e}}{\mathrm{10^{4}\, K}}\right)^{0.175}\\
\times \left(\mathrm{\frac{\rm {\Theta_s}}{arcsec}}\right)^{-1.5}
\left(\mathrm{\frac{D}{kpc}}\right)^{-0.5},
\end{split}
\end{equation}

\begin{equation}
\begin{split}
\label{eqn:eqn2}
\left(\frac{\mathrm{EM}}{\mathrm{pc~cm^{-6}}}\right) = 4.4\times 10^5 
\left( \frac{\nu}{\mathrm{4.9\, GHz}}\right)^{0.1}
\left(\mathrm{\frac{S_\nu}{mJy}}\right)
\left(\frac{\mathrm{T_e}}{\mathrm{10^{4}\, K}}\right)^{0.35} \\
\times \left(\mathrm{\frac{\rm {\Theta_s}}{arcsec}}\right)^{-2},
\end{split}
\end{equation}

 \begin{equation}
 \begin{split}
\label{eqn:eqn3}
\left(\frac{\mathrm{M_{\rm HII}}}{\mathrm{M_{\odot}}}\right) = 3.7\times 10^{-5}
 \left( \frac{\rm \nu}{\mathrm{4.9\, GHz}}\right)^{0.05}
\left(\mathrm{\frac{S_\nu}{mJy}}\right)^{0.5}
\left(\frac{\mathrm{T_e}}{\mathrm{10^{4}\, K}}\right)^{0.175}\\
\times \left(\mathrm{\frac{\rm {\Theta_s}}{arcsec}}\right)^{1.5}
\left(\mathrm{\frac{D}{kpc}}\right)^{2.5},
\end{split}
\end{equation}

\begin{equation}
\begin{split}
\label{eqn:eqn4}
\left(\frac{\rm N'_c}{\rm s^{-1}}\right)~\ge~8.04\times10^{46}\left(\frac{\rm T_e}{\rm K}\right)^{-0.85}\left(\frac{\rm r}{\rm pc}\right)^3\left(\frac{\rm n_e}{\rm cm^{-3}}\right)^2.
\end{split}
\end{equation}

\noindent where $\nu$ is the frequency, S$_{\nu}$ the flux density, T$_{\rm e}$ the electronic temperature, D the distance, r is the radius of the sphere, and $\Theta_{\rm s}$ is its size. The distance and flux density values at 3.6~cm for all continuum sources were taken from Tables \ref{tab:tab3} and \ref{tab:tab4}, respectively, and the size of the sources was calculated using the mean of their two axes. In addition, the ionizing spectral type was determined using \citet{Panagia1973}, considering zero-age main-sequence (ZAMS) objects. 

The physical parameters calculated from the 3.6~cm flux density are listed in Table \ref{tab:tab5}. Most of the calculated parameters for the continuum sources meet the definition of the UC \hii region according to \citet{Wood1989, Kurtz1994}. Although the determination of the physical parameters using the flux density at 3.6~cm is an acceptable approximation, a better characterization requires observations in at least two wavelengths to estimate their spectral index. For this reason, caution must be taken when interpreting these results.

\subsection{Spectral Indices}
\label{sec:sec3.2}
The spectral index provides more reliable information about the nature of the sources. However, to calculate it requires that the sources are detected in at least two wavelengths. The spectral index, $\alpha$ is calculated using a power-law function $S_\nu \propto \nu^{\alpha}$ (being S the flux density at the frequency $\nu$), and its value indicates whether the continuum emission is thermal or non-thermal in nature. For example, at centimeter wavelengths, optically thin H~II regions are associated with a spectral index around -0.1, while optically thick H~II regions have an index $\sim 2$ \citep[e.g][]{Trinidad2003}. Thermal jets, on the other hand, have a spectral index of approximately 0.6 \citep[e.g][and references therein]{Anglada2018}. In contrast, the active magnetosphere of some young low-mass stars has a spectral index ranging from -2 to 2 \citep[e.g][]{Rodriguez2012}, while starburst galaxies have a spectral index ranging from -1.2 to -0.4 \citep[e.g][]{Deeg1993}. Furthermore, the spectral index allows to infer the degree of optical depth of the emission. In the case of thermal emission, its value, together with the morphology, could indicate whether the source is consistent with an \hii region or a thermal jet.

As mentioned, only five sources in the sample were detected at both 3.6 and 1.3~cm. To increase the number of characterized sources, we also used observations at 6~cm from 14 sources reported by \citet{Urquhart2009}.  Out of the 25 sources listed in Table \ref{tab:tab4}, 14 were found to have emissions at 3.6 and 6~cm, while only one source showed emission at both 1.3 and 3.6~cm and four sources were detected at 1.3, 3.6, and 6~cm. 

Because of the 1.3 and 3.6~cm observations have a similar (u,v) coverage and were carried out using the same calibrators with a time difference of about 4.5 months, assuming that their flux density has no significant variations over time, we can estimate a reliable spectral index for the continuum sources detected at these two wavelengths. Although 6~cm observations were carried out over a decade later and with slightly lower angular resolution than those at 1.3 and 3.6~cm, they can still be used to estimate a rough spectral index. As mentioned in Section \ref{sec:sec2}, all data were convolved to have a similar angular resolution (see Table \ref{tab:tab4}). The calculated spectral indices are reported in Table \ref{tab:spectral-index}.

Based on spectral indices and morphology (size, shape, and internal structure, mainly at 3.6~cm), we confirm that the majority of continuum sources could be consistent with \hii regions, five of them associated with optically thick emission, three with optically thin emission, and three with partially optically thin emission. Additionally, we identified four continuum sources with a negative spectral index, which indicates non-thermal emission.

\subsection{H~II Region Models}
\label{sec:sec3.3}
In general, the physical parameters of \hii regions are calculated assuming a homogeneous electron density. However, models that account for specific density distributions, such as the outwardly decreasing density model, are expected to provide a more reliable understanding of the ionized gas physics than the ideal Stromgren sphere model, which does not consider these gradients. One of such model is developed by \citet{Olnon1975}.

Olson's models assume ionized hydrogen gas, circular symmetry for the radius perpendicular along the line of sight, and uniform electron temperature ($T_{e}$). In the Rayleigh-Jeans regime, the total flux density is given by

\begin{equation*}
\label{eqn:eqnA7}
S_{\nu}=\frac{4 \, \pi k \, T_{e} \, \nu^{2}}{c^{2} \, D^{2}}\int_{0}^{\infty} \rho \left[1-e^{-\tau_{\nu}(\rho)}\right] d \rho,  
\end{equation*}
 
\noindent where $\rho$ is the radius perpendicular to the line of sight and $D$ is the distance to the object. 

The optical depth is defined as:

\begin{equation*}
\label{eqn:eqnA8}
\tau_{\nu}(\rho) = f(\nu,T_e) E(\rho) = 8.235 \times 10^{-2} \, T_{e}^{-1.35} \, \nu^{-2.1} E(\rho). 
\end{equation*}

The emission measure can be expressed as

\begin{equation*}
\label{eqn:eqnA9}
 E(\rho)=2 \int_{0}^{\infty} 
n_{e}^{2}(r)dz,
\end{equation*}

\noindent where $r^{2}=\rho^{2}+z^{2}$ and the distance along the line of sight is $z$. With this background and following \citet{Olnon1975}, we explored models with cylindrical, spherical, and Gaussian distributions. 

For the \textit{cylindrical distribution}, we considered a cylinder with radius=$R$ and length=$2 R$, where the electron density $n_{e}$ is constant inside, and zero outside. Therefore: 

\begin{equation}
\label{eqn:eqnA10}
 S_{\nu}=\frac{2 \, \pi k \, T_{e} R^{2} \, \nu ^{2}}{c^{2} \, D^{2}} \left(1-e^{-\tau'}\right). 
\end{equation}

In a similar way, the \textit{spherical distribution} is given by

\begin{equation}
\label{eqn:eqnA11}
 S_{\nu}=\frac{2 \, \pi k \, T_{e} R^{2} \, \nu ^{2}}{c^{2} \, D^{2}} \left[1-\frac{2}{\tau'^{2}}{[1-(\tau'-1)e^{-\tau'}}]\right],
\end{equation}

while the \textit{ Gaussian distribution} is defined by

\begin{equation}
\label{eqn:eqnA12}
 S_{\nu}=\frac{2 \, \pi k \, T_{e} R^{2} \, \nu ^{2}}{c^{2} \, D^{2}} \left[\gamma+ln \, \tau'' + E_{1}(\tau'')\right], 
\end{equation}

\noindent where the \hii region has spherical symmetry, but the electron density distribution is not constant; there is a density gradient with a Gaussian distribution.

In these equations, $R$ is the source radius, $D$ is the distance, $\tau' = 2n^{2}_{0}Rf$, $\tau'' = n^{2}_{0}Rf\sqrt{\pi}$, 
$\gamma$ is the Euler's constant, and E$_1$($\tau$) is the exponential integral, defined as $E_1(x) \equiv \int_{x}^{\infty} (exp^{-t}/t) dt$. We used the least-squares fit method in all these models to find the best values for the radius and density using the minimizing function in Python software. We have

\begin{equation}
\label{eqn:eqnA15}
\chi^{2}=\sum_{i}^{N}\frac{\left[S_{\nu i}^{obs}-S_{\nu 
i}^{mod}(a)\right]^{2}}{ \epsilon_{i}^{2},} 
\end{equation}

\noindent where $S_{\nu i}^{obs}$ is the set of observed data, $S_{\nu i}^{mod}$ is the model, $a$ is the set of parameters in the model to be optimized in the fit, $\epsilon_{i}$ is the estimated uncertainty in the flux density equal to ($SN$/2)($0.15 S_{\nu i}^{obs}$), where $SN$ is the signal to noise ratio.

\section{Discussion}
\label{sec:sec4}

We employed the Olson's models with cylindrical, spherical, and Gaussian distribution to confirm the nature of the \hii regions suggested by the morphology  (see Figures \ref{fig:fig1} and \ref{fig:fig2}), 3.6~cm continuum emission (Table \ref{tab:tab4}), and spectral indices (Table \ref{tab:spectral-index}) for sources detected at two and three wavelengths.  The first two models assume a homogeneous electron density, while the third model uses a density gradient with a Gaussian distribution. For sources detected at two or three bands, we obtained physical parameters using \hii region models with cylindrical (equation \ref{eqn:eqnA10}), spherical (equation \ref{eqn:eqnA11}), and Gaussian (equation \ref{eqn:eqnA12}) geometries. 

Assuming an isothermal ionized gas with a temperature of 10$^4$~K, we minimized equation \ref{eqn:eqnA15} to obtain the best spectral fit for each source. \hii region models were applied to 11 sources from Table \ref{tab:spectral-index} with a spectral index greater than $\sim -0.1$. Of these, eight were detected at two wavelengths, and three were detected at three wavelengths. The resulting best fits for each source are shown in Figure \ref{fig:fig5}, and their respective physical parameters, as determined by the best fit, are listed in Table \ref{tab:models}. However, from our two or three wavelength dataset, we were unable to accurately discriminate between specific models for the symmetry and structure of \hii regions, highlighting the need for additional multi-band observations. In this way, elements such as morphology and inferred substructure from observations can help us in characterizing \hii regions more accurately. For more details on each source, please refer to Appendix \ref{sec:ap-A}.

We present the results in the form of a final catalog, (see Table \ref{tab:tab13}), that summarizes the calculated physical parameters for 20 sources. These were calculated from 3.6~cm emission for sources detected at a single wavelength and from \hii models for sources with two or three observations. Of these sources, 17 show physical parameters consistent with those typical of ultracompact H~II regions (one with cometary morphology) and 3 are compatible with being compact H~II regions. Of the remaining five sources listed in Table \ref{tab:tab5}, 05358-VLA1 has elongated jet-like morphology, while 05305-VLA, 06567-VLA, 21306-VLA, and 21334-VLA have a negative spectral index ($< -0.5$).

\subsection{Detection rate of \hii regions in the sample}
\label{sec:sec4.2}
As mentioned, the sample consists of 106 IRAS sources, 96 of which were observed at 3.6~cm and 52 at 1.3~cm, with 42 of them observed at both wavelengths. The detection rate at 3.6~cm was $\sim25\%$ (25 sources), while at 1.3 cm it was only around 10\% (five sources). There are several reasons that could account for this low detection rate, which will be explored below.

One possible reason for the low detection rate could be the poor sensitivity of the observations, which were made in snapshot mode, with integration times of 5 and 10 minutes at 3.6 and 1.3 cm, respectively. However, even with these integration times, sources with a flux density of $\sim 2$~mJy at 3.6~cm and $\sim 4$~mJy at 1.3~cm could still be detected at $3\sigma$. Thus, this factor can only account for a few cases of non-detection. On the other hand, it is also known that the lifetime of the \hii regions is relatively short, which could also contribute to the low detection rate as well.

We cannot rule out the possibility that the emission measure of potential \hii regions is very large ($>10^9$~pc~cm$^{-6}$), making it optically thick at centimeter/millimeter wavelengths and resulting in a turnover frequency for optically thin emission of around 30~GHz or higher \citep{Kurtz1994}. This would mean that they cannot be detected at 3.6~cm or even at 1.3~cm. On the other hand, \citet{Sewilo2011} observed a small sample of UC and HC~H~II region candidates at several bands and achieved a successful detection rate with flux density ranging from 60 to 350~mJy at 1.3 and 3.6~cm, respectively. Their sources span a range of distances up to 14.0 kpc, which is close to the upper end of the range of distances in our sample. However, the large sample of sources we have explored may include some objects, especially the most compact ones, that could be affected by opacity and become undetectable, particularly at the 6 cm band. Nonetheless, as shown by \citet{Sewilo2011}, this effect is not dominant, at least for the majority of \hii regions observed at wavelengths above a few cm.

\subsection{Non thermal Emission}
\label{sec:sec4.3}

In general, the nature of emission from astronomical sources can be classified as thermal \citep[e.g][]{Olnon1975, Reynolds1986} and non-thermal \citep[e.g][]{Deeg1993}, if the spectral indices are larger than -0.1 or less than -0.5, respectively. We explore some scenarios that could explain the nature of the continuum sources with a negative spectral index.

Negative spectral indices were found in four continuum sources (05305-VLA, 06567-VLA, 21306-VLA, and 21334-VLA), with values ranging from -1.3 to -0.5, which indicate non-thermal emission. Young sources with spectral indices  between $-0.5$ and $-0.1$ have been associated to gyro-synchrotron radiation, produced in strong collisions in radio jets \citep[e.g][]{Trinidad2009} or in the corona of young low-mass stars \citep[e.g][]{Launhardt2022}. Since the sample sources are related to massive star formation regions, the first scenario could be the most likely; however, strong collisions are not expected in \hii regions. Spectral indices as low as $-1.2<\alpha<-0.4$ are only typically found in starburst galaxies \citep[e.g][]{Deeg1993}.

The variability of continuum sources could also explain these negative values of the spectral index, since the observations at 3.6 and 6~cm were carried out about a decade apart. Another possibility is that the spectral index of these sources could be a result  of the emission produce by two or more continuous sources. For example, for the sources IRAS~06567-VLA and IRAS~21306-VLA, there is marginal evidence that the continuous emission is not associated with a single source. In either case, to investigate the nature of these sources, new observations with higher sensitivity and angular resolution at multiple wavelengths are needed.

\section{Conclusions}
\label{sec:sec5}

The \uchii regions are good tracers for places where early-type massive stars form, thus their study and characterization can provide important insights to understand the formation process of high-mass stars. However, due to their short lifetime, the number of known \uchii regions is relatively small. In this context, this paper is intended to increase the number of known \hii regions, mainly ultra-compact, and to provide the basic data that can be used for further detailed investigations. 

We conducted a study on the 1.3 and 3.6~cm continuum emission from a sample of 106 high-luminosity IRAS sources observed with the VLA in its C and B configuration, respectively. 52 sources in the sample were observed at 1.3~cm and 96 at 3.6~cm, with 42 of them observed at both wavelengths. Additionally, we used 6~cm observations reported in the literature of the detected sources. From 3.6~cm observations, we detected 25 sources, while only 5 sources were detected from the 1.3~cm observations. In general, a single radio continuum source was detected toward each IRAS region, although there is marginal evidence of double systems in some regions. We only detected two independent sources in one region.

Using the 3.6~cm emission, we performed an initial characterization of the ionized gas in all detected sources by calculating their traditional physical parameters. For sources that were also detected at 1.3 cm and those with reported 6~cm emission, we determined their spectral index and calculated models of \hii regions with cylindrical, spherical, and Gaussian morphologies. Based on these results, we present a catalog of candidate \hii regions detected in the sample.

\begin{appendices}
\section{Comments on Individual Sources}
\label{sec:ap-A}

\begin{figure*}[!ht]
\includegraphics[trim={0cm 1.9cm 0cm 1.2cm}, clip, width=\columnwidth]{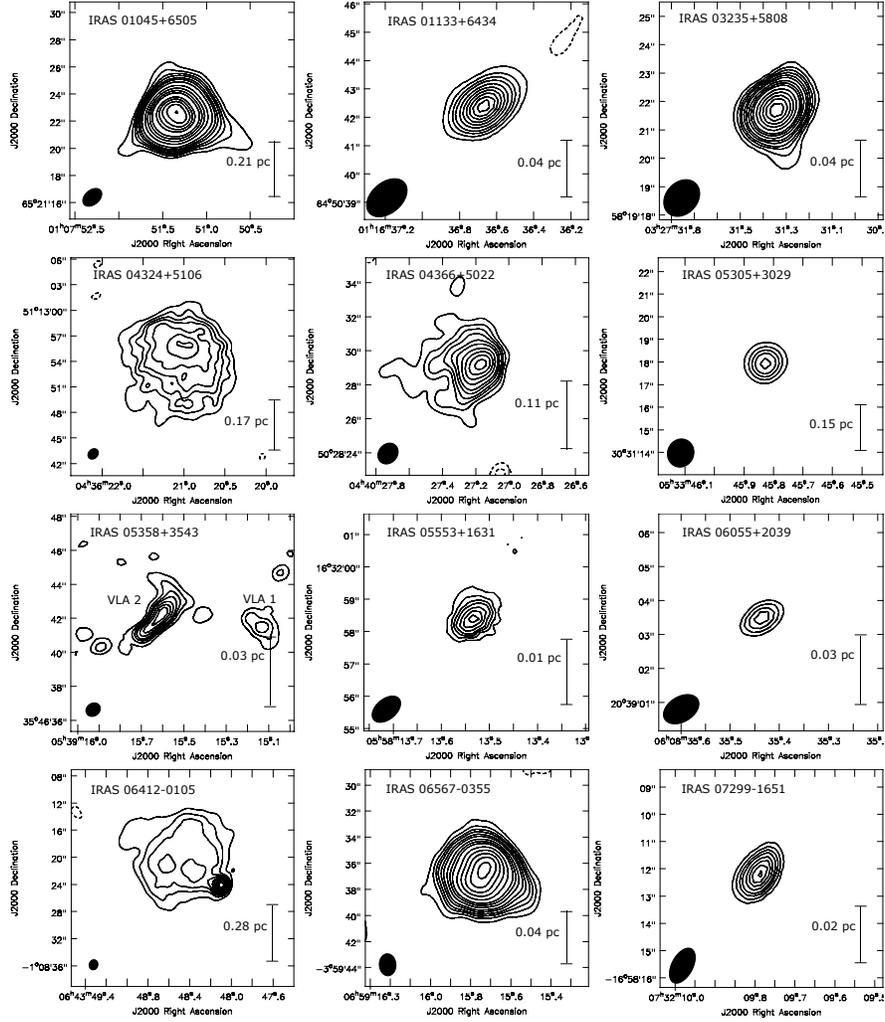}
  \caption{Continuum contour maps of the sources detected at 3.6~cm. The contours for each source are: 
IRAS~01045+6505: -5, 5, 10, 15, 30, 60, 90, 120, 150, 180;
IRAS~01133+6434: -5, -3, 3, 5, 7, 9, 12, 15, 18, 21, 24, 27;
IRAS~03235+5808: -5, -3, 3, 5, 7, 9, 12, 15, 20, 25, 30, 40, 50, 60, 70, 80;
IRAS~04324+5106: -5, -3, 3, 5, 7, 9, 12, 15, 18;
IRAS~04366+5022: -5, -3, 3, 5, 7, 9, 12, 15, 20, 25, 30, 35;
IRAS~05305+3029: -5, -3, 3, 5, 7, 9, 12, 15, 18;
IRAS~05358+3543: -5, -3, 3, 5, 7, 9, 12, 15, 18;
IRAS~05553+1631: -5, -3, 3, 5, 7, 9, 12, 15, 18;
IRAS~06055+2039: -5, -3, 3, 4, 5, 6, 7;
IRAS~06412-0105: -5, -3, 3, 5, 7, 9, 12, 15, 20, 25, 30;
IRAS~06567-0355: -5, -3, 3, 5, 7, 9, 12, 15, 20, 25, 30, 40, 50, 60, 70, 80, and
IRAS~07299-1651: -5, -3, 9, 12, 15, 18, 21 times the respective rms listed in Table \ref{tab:tab4}.
 The beam size is shown at bottom left and given in Table \ref{tab:tab4}.}
 \label{fig:fig1}
\end{figure*}

\begin{figure}[!ht]
\centering
\includegraphics[trim={0cm 1.9cm 0cm 1.2cm}, clip, width=\columnwidth]{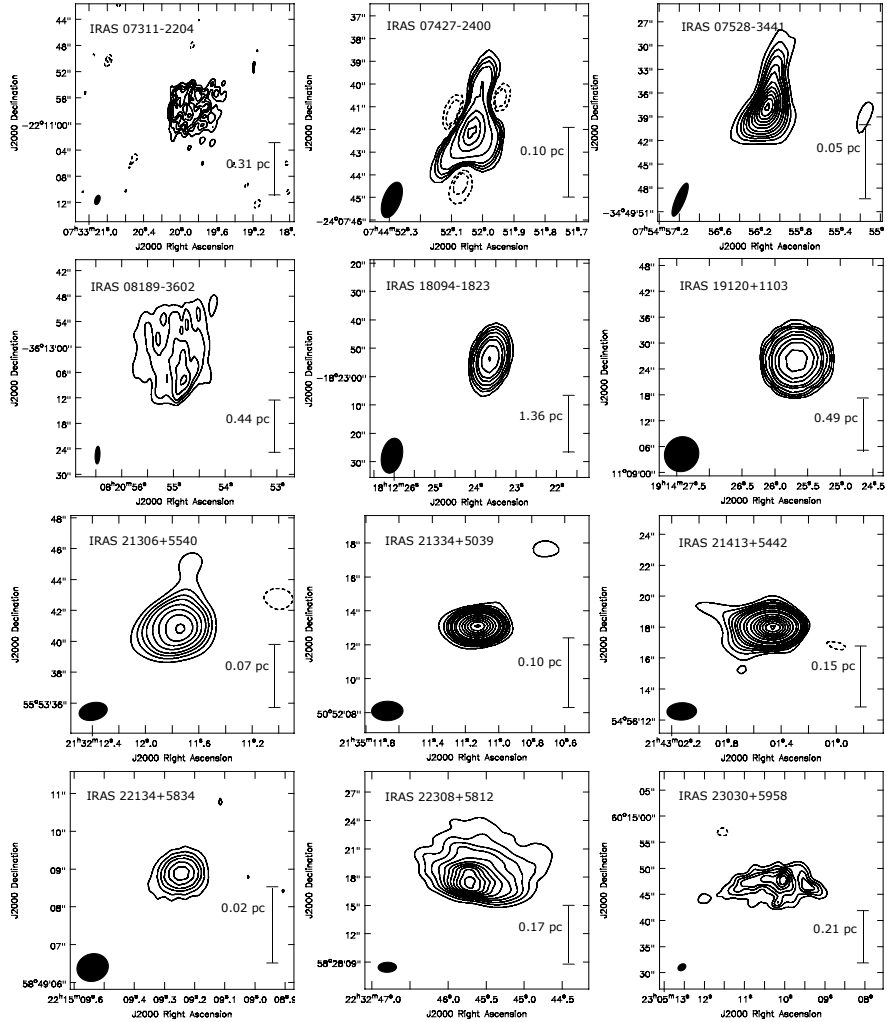}
  \caption{Continuum contour maps of the sources detected at 3.6~cm. The contours for each source are:
IRAS~07311-2204: -5, -3, 3, 5, 7, 9, 12, 15, 18;
IRAS~07427-2400: -5, 5, 10, 20, 40, 60, 80;
IRAS~07528-3441: -5, -3, 3, 5, 7, 9, 12, 15, 20, 25, 30, 35, 40, 45;
IRAS~08189-3602: -4, -3, 3, 4, 5, 7, 9, 11, 13;
IRAS~18094-1823: -5, -3, 3, 5, 7, 10, 15, 20, 30, 45;
IRAS~19120+1103: -5, -3, 3, 5, 7, 10, 15, 20, 30, 40, 60, 80, 100, 150, 200, 250;
IRAS~21306+5540: -5, -3, 3, 5, 7, 9, 12, 15, 18, 21;
IRAS~21334+5039: -5, -3, 3, 5, 7, 9, 12, 15, 20, 25, 35, 45, 55, 65, 75, 85;
IRAS~21413+5442: -5, -3, 3, 5, 7, 9, 12, 15, 20, 30, 40, 60, 80, 100, 120, 140, 160, 180;
IRAS~22134+5834: -5, -3, 3, 5, 7, 9, 12, 15, 18;
IRAS~22308+5812: -5, -3, 3, 5, 7, 9, 11, 13, 15, 17, 19, 21, 23, and
IRAS~23030+5958: -5, -3, 3, 5, 7, 9, 12, 15, 18 times the respective rms listed in Table \ref{tab:tab4}.
 The beam size is shown at bottom left and given in Table \ref{tab:tab4}.}
 \label{fig:fig2}
\end{figure}

\begin{figure}[!ht]
\includegraphics[trim={0cm 7.3cm 0cm 1.2cm}, clip, width=\columnwidth]{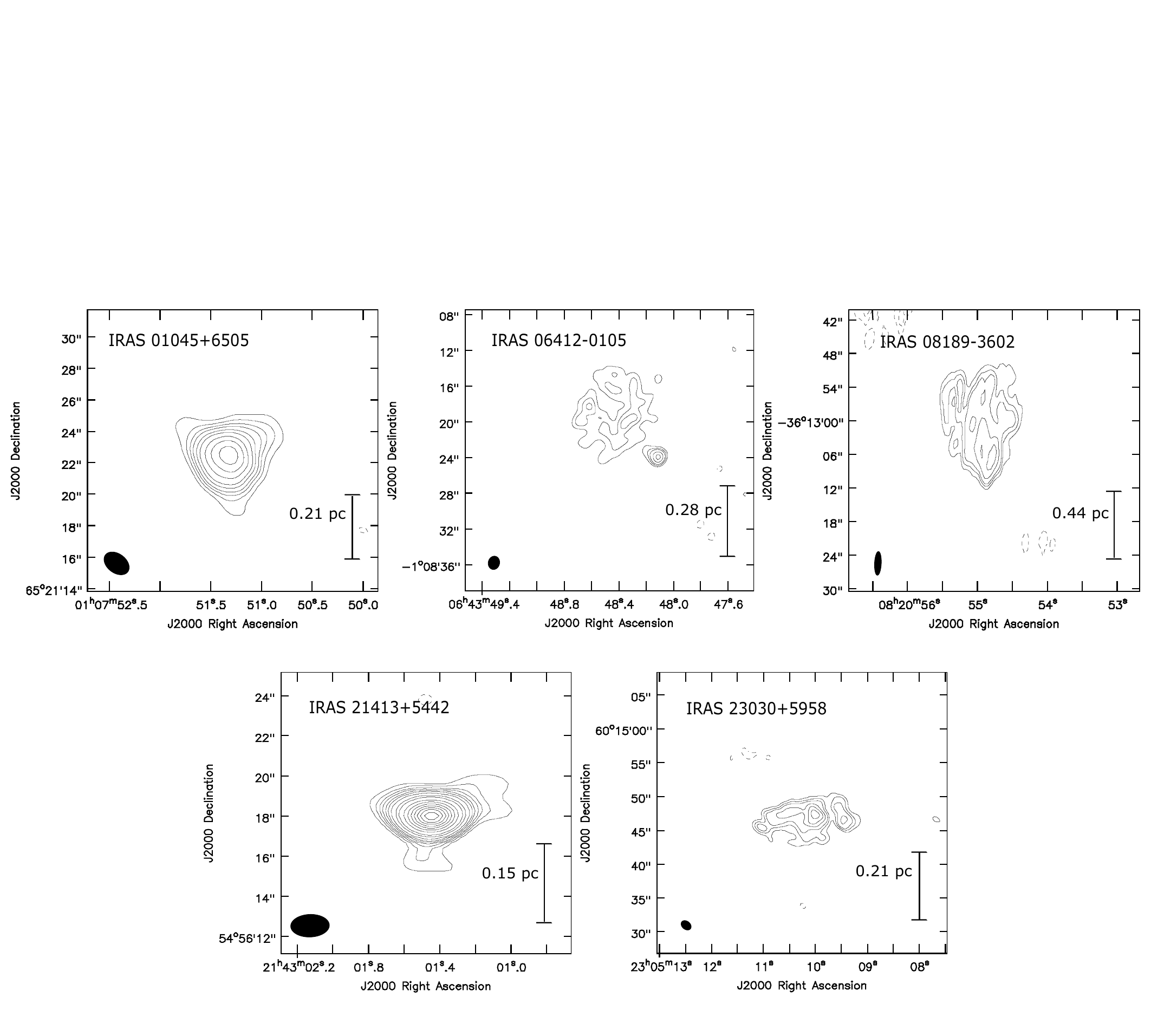}
  \caption{Contour maps of the continuum emission from sources detected at 1.3~cm  The respective contour levels for each source are:
IRAS~01045+6505: -5, -3, 3, 5, 10, 15, 20, 25, 30;
IRAS~06412-0105: -4, -3, 3, 4, 5, 7, 9;
IRAS~08189-3602: -4, -3, 3, 4, 5, 7, 9, 11;
IRAS~21413+5442: -5, -3, 3, 5, 7, 9, 12, 15, 20, 30, 40, 60, 80, 100, 120, 140, 160, 180, and
IRAS~23030+5958: -4, -3, 3, 4, 5, 7, 9, 11 times the respective rms reported in Table \ref{tab:tab4}.
 The beam size is shown at bottom left and given in Table \ref{tab:tab4}.}
 \label{fig:fig3}
\end{figure}

\begin{figure}[!ht]
\centering
\includegraphics[trim={0cm 1.9cm 0cm 1.2cm}, clip, width=\columnwidth]{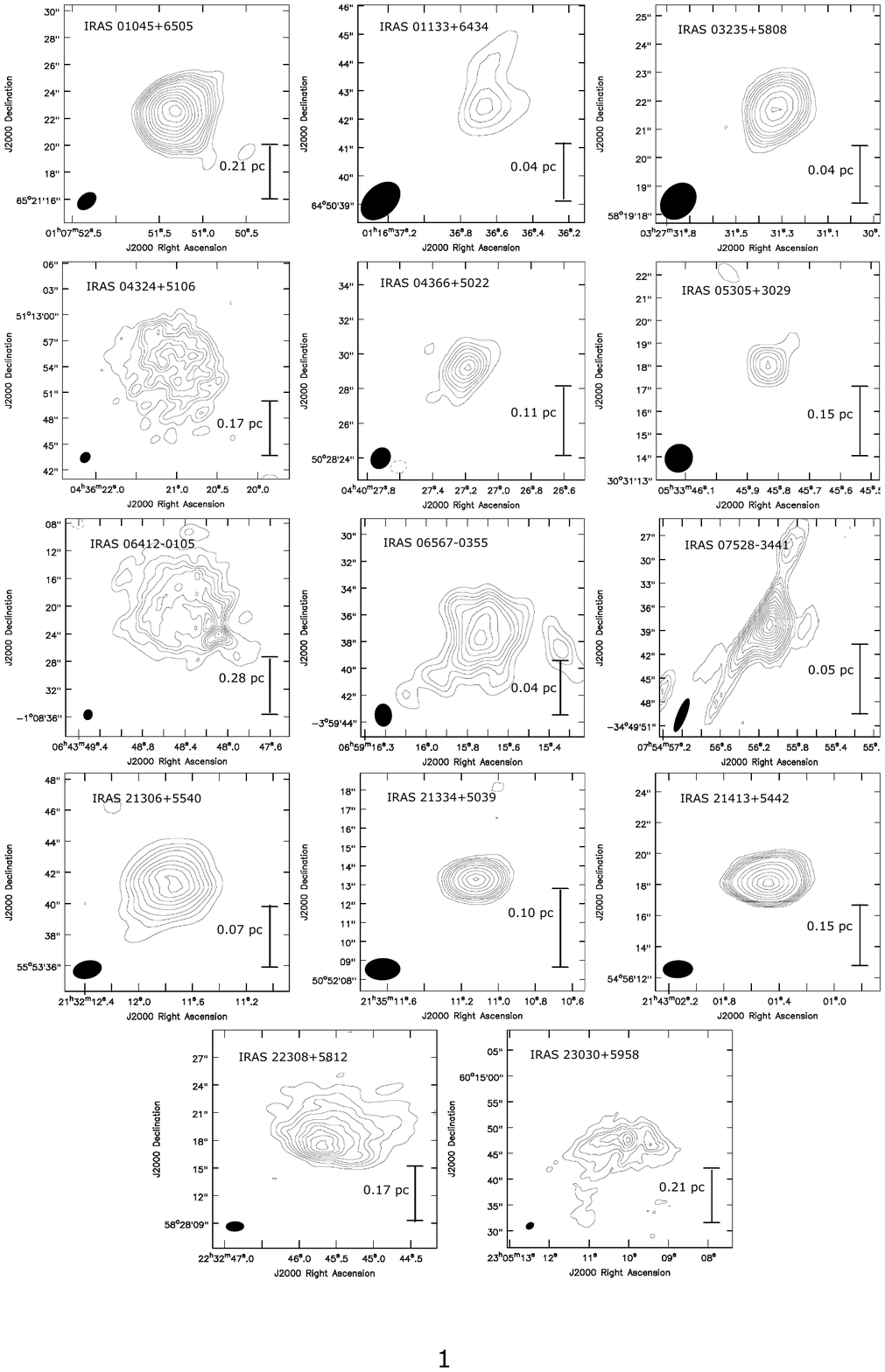}
  \caption{Continuum contour maps of the sources with 6~cm emission (these observations have been reported previously by \citet{Urquhart2009}; see Section \ref{sec:sec2}). The contours for each source are:
Continuum contour maps of the sources detected at 6.0~cm. The contours for each source are:
IRAS~01045+6505:  5, 10, 15, 20, 40, 60, 100, 140, 200;
IRAS~01133+6434: -5, -3, 3, 5, 7, 9, 12, 15, 18;
IRAS~03235+5808: -5, -3, 3, 5, 7, 9, 12, 15, 20, 30, 40, 50;
IRAS~04324+5106: -5, -3, 3, 5, 7, 9, 12, 15, 18;
IRAS~04366+5022: -5, -3, 3, 5, 7, 9, 12, 15, 18, 21;
IRAS~05305+3029: -3, 3, 5, 7, 9, 12, 15, 18;
IRAS~06412-0105: -5, -3, 3, 5, 7, 9, 12, 15, 20, 25, 30;
IRAS~06567-0355: -5, -3, 3, 5, 7, 9, 12, 15, 20, 25, 30;
IRAS~07528-3441: -5, -3, 3, 5, 7, 9, 12, 15, 20, 25, 30, 35, 40, 45, 50, 55;
IRAS~21306+5540: -5, -3, 3, 5, 7, 9, 12, 15, 18, 21, 24;
IRAS~21334+5039: -5, -3, 3, 5, 7, 9, 12, 15, 20, 30, 40, 50, 60;
IRAS~21413+5442: -5, -3, 3, 5, 7, 9, 12, 15, 20, 40, 60, 100, 150, 200, 250;
IRAS~22308+5812: -5, -3, 3, 5, 7, 9, 12, 15, 18, 21, 24, 27, and
IRAS~23030+5958: -5, -3, 3, 5, 7, 9, 12, 15, 18 times the respective rms listed in Table \ref{tab:tab4}.
 The beam size is shown at bottom left and given in Table \ref{tab:tab4}.}
 \label{fig:fig4}
\end{figure}

\begin{figure}[!ht]
\includegraphics[width=\columnwidth]{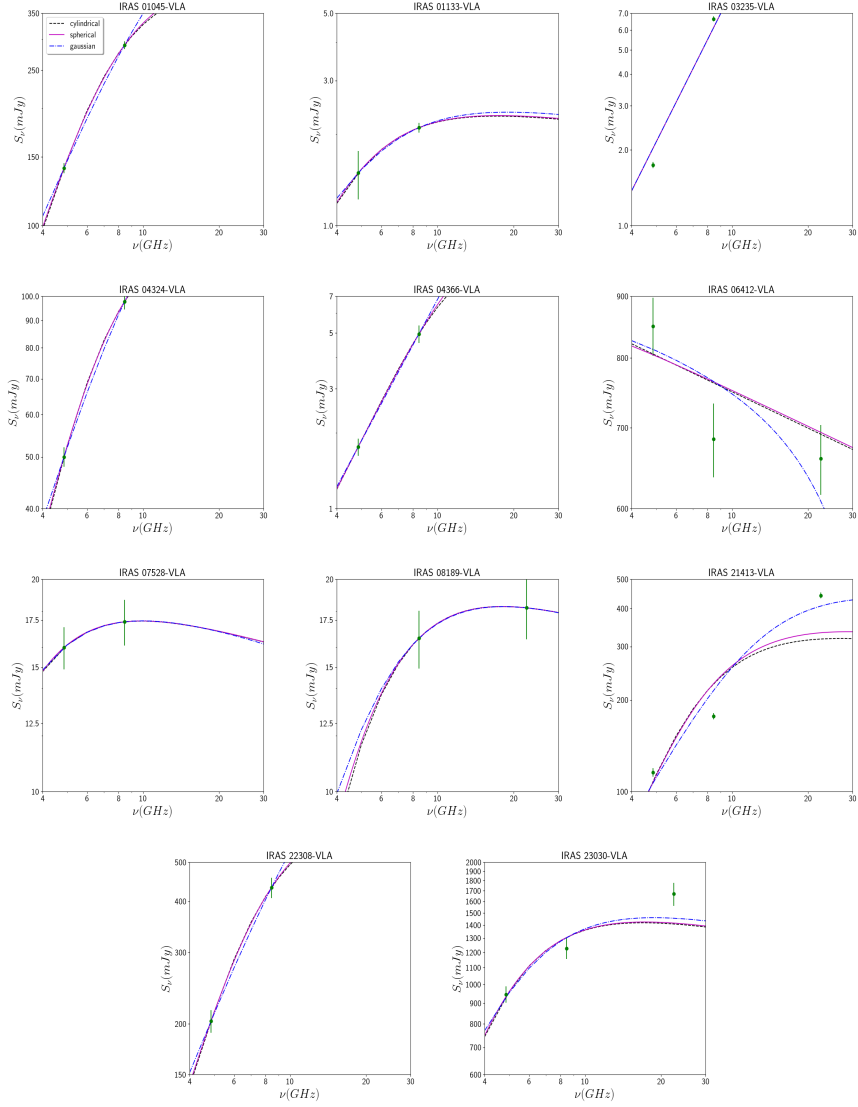}
  \caption{ Fits of the three models of the IRAS sources detected at two or three wavelengths. Dots are the observational data. Dash, continuous, and dash-dot lines are for cylindrical, spherical and Gaussian models respectively. See the text for details.}
 \label{fig:fig5}
\end{figure}

General information is given below for the 25 sources detected at 3.6 cm, as well as the most relevant results of the study carried out in this paper. Table \ref{tab:tab3} lists the luminosity and distance for all sources, while Table \ref{tab:tab5} gives the physical parameters calculated from the 3.6~emission. In addition, Figures \ref{fig:fig1} and \ref{fig:fig2} display the 3.6~cm contour maps, and Figure \ref{fig:fig3} shows the 1.3 cm contour maps of the detected sources in the sample. Contour maps of sources with 6~cm emission are shown in Figure \ref{fig:fig4}. The physical parameters obtained from the models of the \hii regions with cylindrical, spherical and Gaussian symmetry, as well as the best fits to the observational data are given in Table \ref{tab:models} and Figure \ref{fig:fig5}, respectively.\\

{\bf IRAS 01045+6505} is located in the HCS 6236 molecular cloud \citep[][]{Snell2002}. A \uchii region, spatially coincident with a CS molecular clump, and two submillimeter sources has been detected toward it \citep{Mookerjea2007}. We detect 1.3 and 3.6~cm continuum emission toward IRAS~01045+6505. In the field, we observe only one compact source, which is also detected at 6~cm and coincides with the millimeter source 01045-SMM1 and the \uchii region reported by \citet{Mookerjea2007}. A spectral index, $\alpha$ $\sim$ 0.3 was estimated using the flux density of the three wavelengths. However, the flux density at 1.3~cm appears to be very low compared to those obtained at 3.6 and 6~cm, which makes the estimated spectral index unreliable. It is possible that the flux density of the source is variable with the time.

Using only the 3.6 and 6~cm flux densities, we estimated a value of $\alpha$ $\sim$ 1.3. We interpret this spectral index as a partially thick optically \uchii region, which is consistent with the interpretation given by \citet{Mookerjea2007}.  No significant variations are observed in Figure \ref{fig:fig5} between the cylindrical, spherical and Gaussian geometries of the \uchii region; however, based on its morphology at 1.3 and 3.6~cm, the spherical model could be the most suitable.

{\bf IRAS 01133+6434}.  It was previously observed in radio continuum by \citet{Urquhart2009}, finding only one radio source. In our study, we detected a single compact and spherical source in the field at 3.6~cm. However, at 6~cm, this continuum source exhibits two emission peaks, with the strongest one coinciding with the 3.6~cm emission. Using its 3.6 and 6 cm flux density, we estimated a spectral index of $\sim 0.6$, which is consistent with a partially thin optically \hii region. 
The physical parameters obtained from the \hii region's models are given in Table \ref{tab:models}, suggesting that it is an \uchii region sustained by a ZAMS B1 star. We did not find significant differences between the three models applied.

{\bf IRAS 03235+5808}. This source has been little studied. \citet{Urquhart2009} detected a continuum source and NH$_3$ emission in the region. A continuum source is detected at 3.6  and 6~cm, which has a compact and spherical morphology at both wavelengths and coincides spatially with the IRAS center position. Its spectral index ($\alpha \sim 2.4$) and physical parameters suggest that it could be an optically thick \uchii region, which is confirmed by analyzing the \hii region with three symmetries. The central source of the \uchii region is a ZAMS B0 star.

{\bf IRAS 04324+5106}. A radio continuum source and four millimeter sources were detected in the region by \citet{Urquhart2009} and \citet{Klein2005}, respectively. We observed a continuum source at 3.6~cm, that is very extended and shows a cometary morphology. This source is also detected at 6~cm with a similar morphology. Based on its spectral index between 3.6 and 6~cm ($1.2$) and applying the \hii region's models, we find the continuum source is consistent with a compact \hii region with Gaussian symmetry (see Figure \ref{fig:fig5} and Table \ref{tab:models}), which is associated with a ZAMS O9.5 star.

{\bf IRAS 04366+5022}. \citet{Urquhart2009} detect a single continuum source at 6~cm, which has NH$_3$ emission \citep{Urquhart2011}. We detected a continuum source toward IRAS~04366+5022 at wavelengths of 3.6 and 6~cm. This source, at 3.6~cm, shows an irregular morphology with several protuberances, suggesting the presence of more than one embedded continuum source (Figure \ref{fig:fig1}). However, high angular resolution observations will be necessary to confirm this speculation. 

Assuming the continuum emission is produced by a single source, we estimated a spectral index of $\sim 1.9$. This value could be consistent with an optically thick \hii region. Based on the \hii region models, we found that its physical parameters are consistent with this assumption (see Figure \ref{fig:fig5} and Table \ref{tab:models}).

{\bf IRAS 05305+3029}. This source has been poorly studied and no ammonia or other molecular tracer has been detected in the region. A compact continuum source was detected in the field at 3.6~cm, located about $14.3''$ northeast of the IRAS position. Although it was not detected at 1.3~cm,  the continuum source was detected at 6 cm. It shows a compact morphology at 3.6 cm with a protuberance observed at 6 cm, suggesting that the continuum source could be a binary system. In this way, more sensitive observations will be necessary to verify this hypothesis. 

We calculated a spectral index of $\sim -1.2$, which is too negative to be credible. This value could be explained by invoking variability of the continuum source and/or the possibility that it could be a double system. The physical parameters of this continuum source, using its 3.6~cm emission, are consistent with an \uchii region harboring a ZAMS B1 star. However, simultaneous multi-wavelength observations with high angular resolutions are necessary to confirm its nature.

{\bf IRAS 05358+3543} is located toward the star cluster S233 \citep{Yao2000} and has been studied at several wavelengths. Although it is strong at millimeter wavelengths \citep{Beuther2002a}, centimeter continuum emission has not been detected \citep[e.g][]{ Sridharan2002}. In addition, a massive bipolar outflow with a high degree of collimation has been detected \citep{Beuther2002b}.

The 3.6~cm continuum map reveals two sources in the field. The strongest source, IRAS~05358-VLA1, shows an elongated morphology in the northwest-southeast direction, while the second source, IRAS~05358-VLA2, is weaker and more compact. Although IRAS~05358-VLA1 has a jet-like morphology, it was not detected at 1.3~cm or 6~cm, then, its spectral index was not calculated. Moreover, this continuum source is not associated with the millimeter source detected by \citet{Beuther2002a} or the bipolar SiO outflow detected by \citet{Beuther2002b}, whose center is about $75''$ to the southeast. Nonetheless, the elongation of the IRAS~05358-VLA1 is similar to that of the SiO bipolar outflow (northwest-southeast direction). Although IRAS~05358-VLA1 does not seem to be the driver source of the SiO outflow, there may be some relationship. Furthermore, using its 3.6~cm emission, we estimated a mass-loss rate of $\dot{M}$ = 2.71$\times$10$^{-10}$~M$_{\sun}$~yr$^{-1}$, and a momentum rate of $\dot{P}$ = 1.35$\times$10$^{-7}$~M$_{\sun}$~yr$^{-1}$~km~s$^{-1}$, suggesting this source is a thermal jet \citep[e.g.][]{Anglada2018}.

On the other hand, the derived physical parameters of the weak continuum source, IRAS~05358-VLA2, seem to be consistent with a \uchii region hosting a ZAMS B2 star.

{\bf IRAS 05553+1631} is one of the nearest regions in the sample ($\sim1.2$~kpc; \citep{Wouterloot1989}). A millimeter source was detected by \citet{Williams2004} in the region. We detected one compact and spherical continuum source at 3.6~cm, which is offset by approximately $\sim 2.5''$ from the millimeter source. Based on its physical parameters, we suggest that it could be an \uchii region, harboring a ZAMS B3 star.

{\bf IRAS 06055+2039} is located toward S235. Six-millimeter sources and ammonia emission were observed in the region  by \citet{Klein2005}. We detected a weak continuum source toward the IRAS region at 3.6 cm. However, it is shifted about $\sim 49''$ from the IRAS position. Besides, this continuum source is not spatially coincident with any of the other millimeter sources detected by \citet{Klein2005}, the closest one being $1''$ away. Our calculated physical parameters suggest that this compact source is a \hii region with a ZAMS B2 star.

{\bf IRAS 06412--0105} is located toward the WB870 region. A millimeter source was detected in the region by \citet{Klein2005}, which was interpreted as a low-mass dust core embedded into the extended emission. Contour maps at 3.6 and 1.3~cm show a source with cometary morphology that has a size of about $13''\times11''$ at 3.6~cm and is spatially coincident with the IRAS source. A similar morphology is also observed at 6~cm, but the compact source seems to split into a double system. 

Taking into account the compact and extended emission of the continuum source, we estimated a spectral index of $\sim -0.2$, which is interpreted as an optically thin \hii region. In addition, \hii region models indicate that this source is an \uchii region with Gaussian morphology, associated with a ZAMS O7.5 star.

{\bf IRAS 06567--0355}. Both millimetre and IR sources were detected in this region by \citet{Klein2005} and \citet{Zhang1996}, respectively. NH$_3$ (1,1) and (2,2) emission, as well as a bipolar outflow were also reported by \citet[][]{Klein2005} and \citet{Wu1996}, respectively. 

We detected a nearly spherical 3.6~cm continuum source that is shifted about $\sim 290''$ from the IRAS source and is not spatially coincident with the millimeter source. In addition, this continuum source was detected at 6~cm, but with a slightly elongated morphology and showing several protuberances. Using the centimeter emission, we calculate a spectral index of $\sim -0.5$ between 3.6 and 6~cm, which is consistent with non-thermal emission. This spectral index's value could be explain by the presence of variability in the flux density or by the fact the continuum source could be a multiple system. Simultaneous and high angular resolution observations will be necessary to confirm its nature.

{\bf IRAS 07299--1651}. A millimeter and infrared sources, separated by $\sim 2.4''$, were detected by \citet{Klein2005} and \citet{Rosero2019}, respectively. A single continuum source is detected at 3.6~cm in the field, but not at 1.3~cm. It is spatially coincident with the millimeter source, and based on its continuum emission, this source is consistent with an \uchii region maintained by a ZAMS B3 star.

{\bf IRAS 07311--2204} is located toward BRAN 45 region, which has a diameter of $\sim 25'$ and CO emission \citep{May1997}. We detected an extended continuum source at 3.6~cm with angular size $\sim 6''$. This continuum source is spatially coincident with the IRAS source and lies within the BRAN 45 region. Assuming it to be a \hii region and based on its 3.6~cm continuum emission, we found that the ionizing star is a ZAMS B0.5 star.

{\bf IRAS 07427--2400} is a high-mass star-forming region \citep{MacLeod1998}. \citet{Trinidad2011} found a cluster of at least three radio continuum sources, two of which are \uchii regions, while the strongest source is a jet. These sources have also been detected at millimeter wavelengths by \citep{Qiu2009}. 
We detected a continuum source at 3.6 cm; however, we note a peculiar morphology that resembles neither a \hii region nor thermal jets. This fact could be explained due to the low angular resolution of the observations at 3.6~cm, which do not spatially separate the embedded sources detected by \citet{Trinidad2011}.

{\bf IRAS 07528--3441}. Using CS(2--1) and $^{12}$CO observations, \citet{Bronfman1996} found an \uchii region and a molecular outflow toward IRAS 07528--3441. In addition, NH$_3$ (1--1) has been also detected by \citet{Urquhart2011}.  

The 3.6~cm continuum emission shown in Figure \ref{fig:fig2}, seems to be elongated in the north-south direction, as is also observed at 6~cm. A continuum peak is clearly detected and there is evidence of a weaker second peak. However, the second  peak detected at 6~cm is not coincident with the one detected at 3.6~cm. In addition, the continuum emission does not spatially coincide with the \uchii region detected by \citet{Bronfman1996}, which is offset by $\sim 2'$. Based on the spectral index ($\sim 0.15$) and the \hii region models, we found that this continuum source is consistent with an \uchii region harboring a B1 ZAMS star.

{\bf IRAS 08189--3602} was observed as a radio continuum source by \citet[][]{Wouterloot1989}, while \citet{Arnaud2015} found a compact \hii region through mm and sub--mm observations. We detected a continuum source at 3.6 and 1.3~cm with a large angular size of $\sim 20''$ at 3.6~cm. A strong peak at 3.6~cm is observed, but other less defined peaks are also observed. These peaks are not probably associated with other continuum sources, but rather irregularities of extended emission. To investigate the possibility of the secondary emission peaks being associated with compact embedded sources, we made contour maps removing the shorter baselines, both 1.3 and 3.6~cm. However, no additional compact sources were detected. 

Considering the 1.3 and 3.6~cm emission of the source, Figure \ref{fig:fig5} shows that the three \hii region models fit very well for the data. Based on its morphology and physical parameters, we adopt the cylindrical model. However, spherical or Gaussian models can also be consistent. This  \uchii region is separated by $\sim 15''$ from the compact \hii region detected at mm wavelengths by \citet{Arnaud2015}.

{\bf IRAS 18094--1823} (G12.20--0.03) is a high--mass star-forming region \citep{Hill2005} and is not part of the original of the AC295 project. It is located about $\sim$ 4$'$ to the west of the \uchiir~ G12.21--0.10 \citep[e.g.][and references therein]{delaFuente2020a}. It stand out because the presence of low-resolution VLA emission at 3.6 cm (size $\sim$ 20$'$) that coincides with IRAS 18094--1823 in the radio--continuum study for G12.21--0.10 presented by \citet{delaFuente2018}. 
In additon, using~6 cm observations from the CORNISH survey, \citet{Kalcheva2018} suggested that this object is a \uchii region. The 3.6~cm contour map shows a spherical compact source, whose emission is produced by a \uchii region.  We found the ionizing star is a ZAMS B0 star.

{\bf IRAS 19120+1103} (G45.47+0.05) is also not included in the original sample of 104 sources, but VLA low-resolution emission at 3.6~cm was detected in a study by \citet{{delaFuente2020a}}. Its continuum emission is more associated with the \uchii region with extended emission G45.45+0.06 \citep[see][and references therein]{delaFuente2020a}. This source was confirmed as a star-forming region by the detection of H$_2$O and OH maser emission by \citet{Kim2019}. It was classified as an irregular \uchii region base on its 6~cm emission \citep{Wood1989}. The physical parameters and morphology of the 3.6~cm emission are consistent with a compact \hii region, with the central source being a ZAMS O9.5 star.

{\bf IRAS 21306+5540} is located toward S128, which has been studied at radio wavelengths by \citet{Ho1981} and \citet{Fich1986}. Three compact  \hii regions were found by \citet{Ho1981}, labeled as S128A, S128B, and S128N with exciting stars O6, O6, and O9.5, respectively. In addition, an IR and submillimeter source have been detected by \citep{Umana2008}, and the presence of a bipolar outflow in the east-west direction has been reported by \citep{Kim2015}.

We detected a nearly compact continuum source at 3.6 cm, with a protuberance toward the north. This compact source was also detected at 6~cm and coincides with the source S128N detected by \citet{Ho1981}. We estimated a spectral index is about $\sim -1.1$, which is interpreted as non-thermal emission. Such negative spectral indices are generally associated with extra-galactic sources. However, this continuum source is embedded in a star-forming region and has been catalogued as compact \hii region. We could explain this spectral index due to the variability of the source or to the fact that the continuum emission is not associated with a single source (e.g., a protuberance can be observed at 3.6~cm). Its physical parameters are reported in Table \ref{tab:tab5}, assuming it is a \hii region.

{\bf IRAS 21334+5039}. A compact \hii region, with a ZAMS B0 star, was discovered in this region through radio continuum observations by \citet{McCutcheon1991}, which coincides with NH$_3$ emission detected by \citet{Urquhart2011}. In addition, \citet{Obonyo2019} searched non-thermal radio emission toward this region, but the results were negative. 

We detected one continuum source in the field at 3.6~cm, with a compact spherical morphology. A similar morphology was also observed at 6 cm. However, this source has a offset $\sim 60''$  from the compact \hii region detected by \citet[][]{McCutcheon1991}. We determined a spectral index about  $-0.55$, which could indicate a non-thermal nature or variability of the source (observations were carried out with a separation of about ten years). Further simultaneous observations will be necessary to determine its nature.

{\bf IRAS 21413+5442}. Two radio continuum sources have been detected by \citet{Miralles1994} and classified as a compact \hii and a \uchii region, respectively. In addition, IR observations by \citet{Anandarao2008} show the presence  of a stellar cluster. 

We detected a source in the field at 6, 3.6 and 1.3~cm. The morphology of the source is compact, showing slight protrusions in all three wavelengths. It coincides with the IRAS source and one of the IR sources detected by \citet{Anandarao2008}, which was interpreted as a  massive young stellar object. However, this continuum source is offset by about $20'$ from the \hii regions reported by \citet{Miralles1994}. Using the 1.3, 3.6 and 6~cm flux densities, we obtained a spectral index of 0.85, suggesting this source is a partial thick optically \hii region. By modeling this continuum source as \hii region, we find that its physical parameters are consistent with a \uchii region with Gaussian morphology harboring a ZAMS O8 star.

{\bf IRAS 22134+5834}. NH$_3$ and water maser emission were detected toward the IRAS region by \citet{Sunada2007}. We detected a continuum source at 3.6 cm, located about $3.5''$ from the IRAS source. This source shows a compact spherical morphology, and based on its derived physical parameters from the 3.6 cm emission, it could be classified as an \uchii region with a ZAMS B1 star.

{\bf IRAS 22308+5812} is located toward Sh2--138 \citep[][]{Wouterloot1989}. A compact \hii region was studied by \citet{MH2002} and NH$_3$ (1,1) emission was reported by \citet{Urquhart2011}.

We detected a continuum source at 3.6~cm with a cometary-like \hii region morphology, with its emission peak about $2''$ of the IRAS source. A similar morphology was also observed at 6~cm. Based on its spectral index ($\sim 1.4$), physical parameters and morphology observed in Figure \ref{fig:fig2}, we suggest this source is a \uchii region with a ZAMS O7.5 star and Gaussian distribution (see  Figure \ref{fig:fig5}).

{\bf IRAS 23030+5958} is located toward S156 \citep[][]{Lee2012} and it is one of the most luminous regions of the sample. Using low angular resolution observations at 6~cm, \citet{Israel1977} found a group of \hii regions (S156) with at least two O stars and three B stars. 

We detected continuum emission at 3.6 and 1.3 cm, but the morphology does not have a well--defined structure; rather it shows a complicated morphology with at least three continuum peaks detected at both wavelengths (VLA1, VLA2, and VLA3), aligned in the east-west direction. This morphology is also observed at 6 cm. All continuum peaks are contained in the S156A source and interpreted as a \hii region. In general, the morphology of S156A and the other sources detected by \citet{Israel1977} was explained by the quasi-stationary blister type model. Considering that all the emission detected in the field is part of a single \hii region and based on its spectral index information and derived physical parameters, we found that its continuum emission is consistent with a partial optically thin  \uchii region excited by ZAMS O8 stars. 
 
\end{appendices}

\clearpage


\begin{table}[!ht]
\centering
  \setlength{\tabnotewidth}{\columnwidth}
  \small
  \tablecols{6}
\caption{Physical Parameters of \hii regions\tabnotemark{a}.} 
\label{tab:tab1}
\begin{tabular}{lccccr}
\toprule
Type of \hii & Size & EM & n$_{\rm e}$ & M$_{\rm \hii}$ & Reference\tabnotemark{b}  \\
region &  (pc) & (cm$^{-6}$ pc) & (cm$^{-3}$) & (M$_{\odot}$) &  \\
\hline
Hypercompact & $\sim$ 0.003 & $\gtrsim$ 10$^{10}$  & $\gtrsim$ 10$^6$ &  $\sim$ 10$^{-3}$ & 1    \\
Ultracompact & $\lesssim$ 0.1  & $\gtrsim$ 10$^{7}$  & $\gtrsim$ 10$^{4}$  & $\sim$ 10$^{-3}$ & 2,3    \\
Compact & $\lesssim$ 0.5  & $\gtrsim$ 10$^{7}$  & $\gtrsim$ 5$\times$10$^{3}$  & $\sim$ 1 & 4    \\
Ultracompact with & 1--20  & 10$^4$--10$^5$ & $\gtrsim$ 10$^2$-10$^3$ & 5--10$^3$ & 5    \\
Extended Emission & & & &  \\
\bottomrule
\tabnotetext{a}{Adapted from \citet{Kurtz2002, delaFuente2020b}}
\tabnotetext{b}{1.- \citet{Sewilo2008, Sewilo2004}, 2.- \citet{Wood1989}, 3.- \citet{Kurtz1994}, 4.- \citet{Lumsden2013}, 5.- \citep[][and references therein]{delaFuente2020a, delaFuente2020b}.}
\end{tabular}
\end{table}

\begin{table*}[!t]\centering
  \scriptsize
  \newcommand{\DS}{\hspace{6\tabcolsep}} 
  \begin{changemargin}{-2cm}{-2cm}
        \caption{The sample of 106 IRAS sources: 96 observed at 3.6 cm, 10 at 1.3 cm, and 42 observed at both wavelengths. }   
      \label{tab:tab2}
    \setlength{\tabnotewidth}{0.95\linewidth}
    \setlength{\tabcolsep}{1\tabcolsep} \tablecols{9}
    \begin{tabular}{lcclcclcc}
      \toprule
      \textbf{IRAS} & RA (J2000)  & DEC (J2000) & \textbf{IRAS} & RA (J2000) & DEC (J2000) & \textbf{IRAS} & RA (J2000) & DEC (J2000)  \\
      \textbf{Source} & (h:m:s) & ($\circ$: $\prime$: $\prime\prime$ ) & \textbf{Source} & (h:m:s) & ($\circ$: $\prime$: $\prime\prime$ ) & \textbf{Source} & (h:m:s) & ($\circ$: $\prime$: $\prime\prime$ ) \\
      \midrule
\textbf{00117+6412*} &  00:14:27.72 & 64:28:46.3  &  \textbf{05554+2013}    &  05:58:24.56  &  20:13:57.5     & \textbf{07528-3441}  &  07:54:49.97 & -34:49:45.9    \\
00338+6312** &   00:36:47.51 & 63:29:02.1  &  06055+2039*   &  06:08:32.82  & 20:39:16.2   &  07530-3436**    &   07:54:56.18  &  -34:49:38.3   \\ 
00412+6638*  &   00:44:15.23 & 66:54:40.6  &  06073+1249*   &  06:10:12.43  &  12:48:45.5   &  08007-2829     &   08:02:46.36  &  -28:25:47.4    \\
00468+6508 	 &   00:49:55.82 & 65:43:38.7  &  06084+1727    &  06:11:24.52  &  17:26:26.5   &  08008-3423     &   08:02:42.30  &  -34:31:46.8    \\
00468+6527** &   00:49:55.82 & 65:43:38.7  &  06089+1727**  &  06:11:44.41  &  17:26:05.1   &  08088-3554*    &   08:10:43.49  &  -36:03:29.8    \\
00556+6048 	 &   00:58:40.13 & 61:04:44.0  &  06103+1523    &  06:13:18.21  &  15:23:16.1   &  08140-3556     &   08:15:58.98  &  -36:08:20.0     \\
00578+6233 	 &   01:00:55.81 & 62:49:28.5  &  06104+1524A** &  06:13:21.32  &  15:23:56.9   &  08159-3543     &   08:17:52.89  &  -35:52:49.9      \\
01045+6505*  &   01:07:50.70 & 65:21:21.4  &  06105+1756*   &  06:13:28.33  &  17:55:29.5   &  08189-3602*    &   08:20:47.86  &  -36:12:34.4     \\
01133+6434*  &   01:16:37.39 & 64:50:38.8  &  06114+1745*   &  06:14:23.69  &  17:44:36.5   &  08212-4146     &   08:23:02.96  &  -41:55:48.5     \\
02044+6031*  &   02:08:05.05 & 60:45:56.7  &  06155+2319A   &  06:18:35.15  &  23:18:11.4   &  08245-4038*    &   08:26:17.70  &  -40:48:35.1      \\
02395+6244 	 &   02:43:28.72 & 62:57:05.3  &  06208+0957*   &  06:23:34.41  &  09:56:22.1   &  08274-4111     &   08:29:13.94  &  -41:10:44.4   \\
02437+6145*  &   02:47:40.43 & 61:58:26.3  &  06306+0437*   &  06:33:16.36  &  04:34:56.8   &  18094--1823\tabnotemark{a} & 18:12:23.63 & -18:22:53.7   \\ 
02455+6034 	 &   02:49:23.23 & 60:47:01.2  &  06331+1102    &  06:35:56.01  &  11:00:17.5   &  19120+1103\tabnotemark{b}  & 19:14:25.67 & 11:09:26.0     \\
02461+6147 	 &   02:50:08.11 & 61:59:47.1  &  06337+1051    &  06:36:29.48  &  10:49:05.1   &  21074+4949     &   21:09:08.09  &  50:01:59.8     \\
03233+5809** &   03:27:22.33 & 58:19:45.8  &  06381+1039    &  06:40:58.00  &  10:36:48.8   &  21080+4950     &   21:09:42.83  &  50:08:29.5       \\
03235+5808*  &   03:27:31.15 & 58:19:21.3  &  06412-0105*   &  06:43:44.97  &  -01:08:06.7  &  21202+5157*    &   21:21:53.18  &  52:10:43.6  \\
04034+5107 	 &   04:07:11.93 & 51:24:44.7  &  06426+0025    &  06:45:15.50  &  00:22:25.9   &  21290+5535     &   21:30:38.70  &  55:48:59.6    \\
04324+5102 	 &   04:36:16.08 & 51:08:12.8  &  06446+0029    &  06:47:12.87  &  00:26:06.5   &  21306+4927**   &   21:05:15.62  &  49:40:01.2  \\
04324+5106*  &   04:36:19.70 & 51:12:44.6  &  06501+0143    &  06:52:45.57  &  01:40:14.9   &  21306+5540*    &   21:32:11.56  &  55:53:23.7  \\  
04366+5022*  &   04:40:26.12 & 50:28:24.7  &  06547-0109A   &  06:57:16.69  &  -01:13:39.5  &  21334+5039*    &   21:35:09.18  &  50:53:09.2  \\
04547+4753*  &   04:58:29.66 & 47:58:27.6  &  06567-0355*   &  06:59:15.76  &  -03:59:39.0  &  21334+5329     &   21:35:05.86  &  53:43:01.2   \\
04579+4703 	 &   05:01:39.74 & 47:07:23.1  &  06570-0401    &  06:59:30.95  &  -04:05:35.1  &  21407+5441*    &   21:42:23.68  &  54:55:06.7     \\ 
05100+3723 	 &   05:13:25.43 & 37:27:04.5  &  07024-1102    &  07:04:45.65  &  -11:07:14.5  &  21413+5442*    &   21:43:01.36  &  54:56:16.3     \\ 
05271+3059 	 &   05:30:21.22 & 31:01:27.2  &  07069-1045    &  07:04:45.65  &  -11:07:14.5  &  22134+5834*    &   22:15:09.08  &  58:49:09.3     \\
05274+3345*  &   05:30:45.62 & 33:47:51.6  &  07061-0414*   &  07:08:38.75  &  -04:19:07.5  &  22308+5812*    &   22 32 46.01  &  58 28 21.8      \\
05281+3412   &   05:31:26.60 & 34:14:57.7  &  07207-1435    &  07:23:01.28  &  -14:41:32.5  &  22475+5939*    &   22:49:29.47  &  59:54:56.6    \\ 
05305+3029*  &   05:33:44.81 & 30:31:04.5  &  07295-1915**  &  07:33:10.45  &  -19:28:42.9  &  22502+5944**   &   22:51:59.86  &  59:59:16.9      \\
05334+3149 	 &   05:36:41.08 & 31:51:13.8  &  07298-1919    &  07:32:02.46  &  -19:26:02.3  &  22506+5944*    &   22:52:38.63  &  60:00:55.8   \\
05358+3543*  &   05:39:10.39 & 35:45:19.2  &  07299-1651*   &  07:32:10.00  &  -16:58:14.7  &  22539+5758     &   22:56:00.01  &  58:14:45.9    \\
05361+3539 	 &   05:39:27.66 & 35:40:43.0  &  07333-1838    &  07:35:34.31  &  -18:45:32.5  &  22551+6139     &   22:57:11.23  &  61:56:03.4    \\ 
05375+3536   &   05:40:52.52 & 35:38:23.8  &  07334-1842    &  07:35:40.95  &  -18:48:59.0  &  22570+5912*    &   22:59:06.50  &  59:28:27.7    \\
05375+3540*  &   05:40:53.64 & 35:42:15.7  &  07422-2001    &  07:44:27.85  &  -20:08:31.9  &  23030+5958*    &   23:05:10.62  &  60:14:40.4    \\  
05490+2658 	 &   05:52:12.93 & 26:59:32.9  &  07427-2400*   &  07:44:51.90  &  -24:07:40.6  &  23033+5951     &   23:05:25.16  &  60:08:11.6    \\
05480+2545*  &   05:51:10.75 & 25:46:14.3  &  07311-2204*   &  07:33:20.24  &  -22:10:57.7  &  23139+5939     &   23:16:09.32  &  59:55:22.8    \\ 
05553+1631*  &   05:58:13.87 & 16:32:00.1  &  07434-2044    &  07:45:35.47  &  -20:51:38.6  &  23151+5912     &   23:17:21.09  &  59:28:48.8    \\
             &               &             &                &               &               &  23545+6508**   &   23:57:05.23  &  65:25:10.8   \\
      \bottomrule
\tabnotetext{*}{Sources observed at 3.6 and 1.3~cm at high resolution.} 
\tabnotetext{**}{Sources observed at 1.3~cm at high resolution.} 
\tabnotetext{a}{Source refereed as 18094--G12.20.  Low resolution at 3.6 cm observation only. See text for details.} 
\tabnotetext{b}{Source refereed as 19120--G45.47. Arguable designation: the IRAS source is more related with G45.45+0.06. Low resolution at 3.6~cm only.  See text for details.}
    \end{tabular}
  \end{changemargin}
\end{table*}

\begin{table}[!ht]
\centering
  \setlength{\tabnotewidth}{\columnwidth}
  \small
  \tablecols{4}
\caption{Phase calibrator's observational parameters.} 
\label{tab:tab_2}
\begin{tabular}{lccc}
\toprule
Calibrator & RA (J2000) & DEC (J2000)  & 3.6 Bootstrapped Flux Density \\
 &  (h:m:s) & ($\circ$: $\prime$: $\prime\prime$ ) & (Jy)   \\
\hline
2023+544  & 20h23m55.844s &   54$^{\circ}$27$'$35.83$''$  &  1.12$\pm$0.01 \\
2230+697  & 22h30m36.470s &   69$^{\circ}$46$'$28.08$''$  &  0.44$\pm$0.01 \\
0228+673  & 02h28m50.051s &   67$^{\circ}$21$'$03.03$''$  &  0.78$\pm$0.02 \\
0359+509  & 03h59m29.747s &   50$^{\circ}$57$'$50.16$''$  &  1.47$\pm$0.03 \\
0555+398  & 05h55m30.806s &   39$^{\circ}$48$'$49.17$''$  &  4.86$\pm$0.08 \\
0530+135  & 05h30m56.417s &   13$^{\circ}$31$'$55.15$''$  &  1.40$\pm$0.02 \\
0700+171  & 07h00m01.525s &   17$^{\circ}$09$'$21.70$''$  &  1.11$\pm$0.01 \\
0725--009 & 07h25m50.640s &   00$^{\circ}$54$'$56.54$''$  &  0.95$\pm$0.01 \\
0730--116 & 07h30m19.112s & --11$^{\circ}$41$'$12.60$''$  &  4.63$\pm$0.05 \\
0828--375 & 08h28m04.780s & --37$^{\circ}$31$'$06.28$''$  &  1.07$\pm$0.01 \\
\bottomrule
\end{tabular}
\end{table}

\begin{table}[htp]\centering
  \setlength{\tabnotewidth}{\columnwidth}
\scriptsize
  \tablecols{6}
\caption{Sources detected at 3.6~cm  }
\label{tab:tab3}
\begin{tabular}{cccccc}\toprule
IRAS\tabnotemark{a}  & VLA 3.6~cm   & RA (J2000)  & Dec (J2000) & Distance\tabnotemark{b} & L$_{\rm FIR}$\tabnotemark{b} \\
Source  &  Source    & (h:m:s) & ($\circ$: $\prime$: $\prime\prime$ ) & (kpc) & (10$^4$L$_{\odot}$)  \\  \midrule
\, 01045+6505                                & 01045--VLA    & 01:07:51.34 & 65:21:22.4 & 10.7\tabnotemark{1} &  8.00\tabnotemark{17}   \\
\, 01133+6434                          & 01133--VLA    & 01:16:36.67 & 64:50:42.4 & 4.1\tabnotemark{2}  &  0.84\tabnotemark{2}   \\
\, 03235+5808                            & 03235--VLA    & 03:27:31.34 & 58:19:21.7 & 4.2\tabnotemark{2}  &  1.30\tabnotemark{2}   \\
\, 04324+5106                                 & 04324--VLA    & 04:36:21.03 & 51:12:54.7 & 5.8\tabnotemark{3}  &  6.00\tabnotemark{3}   \\
\, 04366+5022                                 & 04366--VLA    & 04:40:27.20 & 50:28:29.2 & 5.9\tabnotemark{3}  &  3.00\tabnotemark{3}   \\
\, 05305+3029                      & 05305--VLA    & 05:33:45.83 & 30:31:18.0 & 10.4\tabnotemark{4} &  0.60\tabnotemark{4}  \\
\, 05358+3543                          & 05358--VLA1   & 05:39:15.62 & 35:46:42.1 & 1.8\tabnotemark{5}  &  0.66\tabnotemark{5}   \\
                                                  & 05358--VLA2   & 05:39:15.13 & 35:46:41.6 & 1.8\tabnotemark{5}  &  0.66\tabnotemark{5}  \\
\, 05553+1631                                & 05553--VLA    & 05:58:13.53 & 16:31:58.4 & 1.2\tabnotemark{3}  &  0.20\tabnotemark{3}   \\
\, 06055+2039                            & 06055--VLA    & 06:08:35.44 & 20:39:03.5 & 2.9\tabnotemark{3}  &  3.00\tabnotemark{3}   \\
\, 06412--0105                           & 06412--VLA    & 06:43:48.42 & -01:08:20.5 & 7.1\tabnotemark{3}  & 9.00\tabnotemark{3}    \\
\, 06567--0355                             & 06567--VLA    & 06:59:15.74 & -03:59:36.8 &  2.3\tabnotemark{6} & 1.80\tabnotemark{18}   \\
\, 07299--1651                               & 07299--VLA    & 07:32:09.79 & -16:58:12.2 & 1.4\tabnotemark{3}  & 0.70\tabnotemark{3}  \\  
\, 07311--2204                        & 07311--VLA    & 07:33:19.92 & -22:10:57.5 & 8.0\tabnotemark{7}  & 20.00\tabnotemark{7}  \\
\, 07427--2400                           & 07427--VLA    & 07:44:52.03 & -24:07:42.1 & 6.9\tabnotemark{3}  & 50.10\tabnotemark{19} \\  
\, 07528--3441                             & 07528--VLA    & 07:54:56.12 & -34:49:37.8 & 1.2\tabnotemark{8}  & 20.00\tabnotemark{8}  \\
\, 08189--3602                        & 08189--VLA    & 08:20:54.92 & -36:13:02.5 & 7.6\tabnotemark{3}  & 30.00\tabnotemark{20}  \\
\, 18094--1823\tabnotemark{c}       & 18094--G12.20 & 18:12:23.63 & -18:22:53.7 & 14.0\tabnotemark{13} & 86.80\tabnotemark{14} \\
\, 19120+1103\tabnotemark{d}         & 19120--G45.47 & 19:14:25.67 & 11:09:26.0 & 8.4\tabnotemark{15}   &  49.2\tabnotemark{16}   \\
\, 21306+5540                              & 21306--VLA    & 21:32:11.76 & 55:53:40.9 & 3.7\tabnotemark{9}   & 1.10\tabnotemark{3}    \\
\, 21334+5039                          & 21334--VLA    & 21:35:11.13 & 50:52:13.1 & 5.0\tabnotemark{10}  & 2.10\tabnotemark{10}   \\
\, 21413+5442                                & 21413--VLA    & 21:43:01.47 & 54:56:18.0 & 7.9\tabnotemark{11}  & 1.45\tabnotemark{11}    \\
\, 22134+5834                 &   22134--VLA     & 22:15:09.25 & 58:49:08.9 & 2.3\tabnotemark{3}   & 1.34\tabnotemark{3}    \\
\, 22308+5812                              & 22308--VLA    & 22:32:45.62 & 58:28:18.2 & 5.7\tabnotemark{3}   & 9.00\tabnotemark{3}    \\
\, 23030+5958                               & 23030--VLA    & 23:05:10.20 & 60:14:47.2 & 4.4\tabnotemark{12}  & 10.00\tabnotemark{3}   \\
\bottomrule
\tabnotetext{a}{The observed source does not necessary coincides with the IRAS source.}
\tabnotetext{b}{The distance and the FIR luminosity are from the IRAS region, and not necessary corresponds to the observed sources at 3.6 cm. Values taken from: 1.- \citet{Rudolph1996}, 2.- \citet{Maud2015}, 3.- \citet{Wouterloot1989}, 4.- \citet{Lumsden2013}, 5.- \citet{Lu2014}, 6.- \citet{Tapia1997}, 7.- \citet{May1997}, 8.- \citet{Preite-Martinez1988}, 9.- \citet{Kim2015}, 10.- \citet{McCutcheon1991}, 11.- \citet{Navarrete2015}, 12.- \citet{Lee2012}, 13.- \citet{Hill2005}, 14.- We assume the IRAS FIR luminosity of G12.21--0.10 \citep{delaFuente2018,delaFuente2020a}, 15.- \citet{Wu2019}, 16.- We assume the IRAS FIR luminosity of G45.45+0.06 \citep{delaFuente2020a}, 17.- \citet{Snell2002}, 18.- \citet{Klein2005}, 19.- \citet{MacLeod1998}, 20.- \citet{Arnaud2015}.} 
\tabnotetext{c}{This source was not included in the original sample of 94 sources (see Table \ref{tab:tab_2}).}
\tabnotetext{d}{This source was not included in the original sample of 94 sources (see Table \ref{tab:tab_2}). The nearest IRAS source is 19120+1103, but this coincide in position with the \uchii region with extended emission G45.455+0.058 or G45.45+0.06 \citep{delaFuente2020a}. See text for discussion. The distance is adopted from \citet{Wu2019}.}
\end{tabular}
\end{table}

\begin{table}[!htp]
\small
\centering
 \setlength{\tabnotewidth}{\columnwidth}
  \tablecols{7}
\caption{Observational parameters of the sources detected at 1.3, 3.6, and 6.0 cm}
\label{tab:tab4}
\scriptsize
\begin{tabular}{lcccccc}\toprule
\, VLA 3.6~cm    & $\lambda$ & S$_{\nu}$         & Beam Size          & PA    & RMS Noise         & Size               \\
\,  Source       & (cm)      & (mJy)             & ( '' $\times$ '' ) & (deg) & (mJy beam$^{-1}$) & ( '' $\times$ '' )  \\  \midrule
\, 01045--VLA   & 6.0       & 140.4$\pm$4.2   & 1.57$\times$1.05   & 134    & 0.25              & 3.13$\times$2.91    \\
\,               & 3.6       & 289.8$\pm$6.7   & 1.57$\times$1.05   & 120    & 0.41              & 3.10$\times$2.92    \\
\,               & 1.3       & 100.5$\pm$5.4   & 1.57$\times$1.05   & 27    & 0.67              & 3.12$\times$3.00    \\
\, 01133--VLA   & 6.0       & 1.5$\pm$0.3     & 1.61$\times$1.07   & 168   & 0.06              & 3.11$\times$1.73   \\
\,               & 3.6       & 2.1$\pm$0.1     & 1.61$\times$1.07   & 129   & 0.06              & 1.75$\times$1.15   \\
\, 03235--VLA   & 6.0       & 1.7$\pm$0.1     & 1.41$\times$1.10   & 146   & 0.03              & 1.50$\times$1.13   \\
\,               & 3.6       & 6.6$\pm$0.2     & 1.41$\times$1.10   & 144   & 0.07              & 1.48$\times$1.16   \\
\, 04324--VLA   & 6.0       & 50.0$\pm$2.1    & 1.34$\times$1.08   & 47    & 0.06              & 8.82$\times$7.27   \\
\,               & 3.6       & 97.7$\pm$3.3    & 1.34$\times$1.08   & 46    & 0.12              & 8.85$\times$8.31   \\
\, 04366--VLA   & 6.0       & 1.8$\pm$0.1     & 1.30$\times$1.08   & 141   & 0.03              & 2.41$\times$1.73 \\
\,               & 3.6       & 5.0$\pm$0.4     & 1.30$\times$1.08   & 158   & 0.03              & 2.72$\times$2.27   \\
\, 05305--VLA   & 6.0       & 0.4$\pm$0.1     & 1.26$\times$1.15   & 146   & 0.02              & 1.32$\times$1.15 \\
\,               & 3.6       & 0.21$\pm$0.01     & 1.26$\times$1.15   & 120   & 0.02              & 1.30$\times$1.21   \\
\, 05358--VLA1   & 3.6       & 1.8$\pm$ 0.2    & 0.88$\times$0.74   & 139   & 0.02              & 3.04$\times$1.05   \\
\, 05358--VLA2   & 3.6       & 0.8$\pm$ 0.1    & 0.88$\times$0.74   & 30   & 0.02              & 2.35$\times$1.49   \\
\, 05553--VLA   & 3.6       & 0.8$\pm$0.1     & 1.00$\times$0.62   & 135   & 0.03              & 1.11$\times$0.82   \\
\, 06055--VLA   & 3.6       & 0.8$\pm$0.1     & 1.11$\times$0.74   & 124   & 0.09              & 1.30$\times$0.82   \\
\, 06412--VLA   & 6.0       & 850.0$\pm$47.0  & 1.51$\times$1.25   & 57    & 0.67              & 13.85$\times$12.17 \\
\,               & 3.6       & 685.0$\pm$48.0  & 1.51$\times$1.25   & 62    & 0.90              & 12.97$\times$10.74   \\
\,               & 1.3       & 660.0$\pm$44.0  & 1.51$\times$1.25   & 146   & 1.90              & 12.59$\times$12.36   \\
\, 06567--VLA   & 6.0       & 49.3$\pm$3.2    & 1.69$\times$1.24   & 161   & 0.21              & 4.58$\times$3.65 \\
\,               & 3.6       & 37.0$\pm$1.1    & 1.69$\times$1.24   & 166   & 0.07              & 3.68$\times$3.53   \\
\, 07299--VLA   & 3.6       & 0.26$\pm$0.01     & 1.40$\times$0.76   & 148   & 0.01              & 1.44$\times$0.86   \\  
\, 07311--VLA   & 3.6       & 4.0$\pm$0.3     & 1.50$\times$0.76   & 165   & 0.01              & 6.26$\times$5.28   \\
\, 07427--VLA   & 3.6       & 2.3$\pm$0.2     & 1.65$\times$0.77   & 160   & 0.02              & 1.81$\times$0.95   \\  
\, 07528--VLA   & 6.0       & 16.0$\pm$1.1    & 4.52$\times$1.14   & 153   & 0.08              & 5.57$\times$3.30   \\
\,               & 3.6       & 17.4$\pm$1.3    & 4.52$\times$1.14   & 163   & 0.12              & 6.08$\times$2.85   \\
\, 08189--VLA   & 3.6       & 16.5$\pm$1.6 & 4.24$\times$1.17   & 176   & 2.72              & 26.66$\times$16.20  \\
\,               & 1.3       & 18.2$\pm$1.8  & 4.24$\times$1.17   & 179   & 2.45              & 19.43$\times$10.77  \\
\, 18094--G12.20 & 3.6       & 7.2$\pm$0.1     & 12.54$\times$7.26  & 167   & 0.15              & 12.84$\times$7.31   \\
\, 19120--G45.47 & 3.6       & 112.4$\pm$1.7   & 8.23$\times$7.64   & 159   & 1.34              & 8.34$\times$7.63   \\
\, 21306--VLA   & 6.0       & 74.0$\pm$3.3    & 1.81$\times$1.14   & 116   & 0.46              & 3.87$\times$3.37   \\
\,               & 3.6       & 39.9$\pm$2.6    & 1.81$\times$1.14   & 128   & 0.40              & 3.35$\times$2.76   \\
\, 21334--VLA   & 6.0       & 7.7$\pm$0.1     & 1.80$\times$1.15   & 93    & 0.12              & 1.82$\times$1.21   \\
\,               & 3.6       & 5.7$\pm$0.1     & 1.80$\times$1.15   & 93    & 0.06              & 1.84$\times$1.17   \\
\, 21413--VLA   & 6.0       & 115.7$\pm$3.7   & 1.87$\times$1.13   & 93    & 0.29              & 2.18$\times$1.40   \\
\,               & 3.6       & 177.1$\pm$4.4   & 1.87$\times$1.13   & 91    & 0.73              & 2.13$\times$1.32   \\
\,               & 1.3       & 441.8$\pm$9.8   & 1.87$\times$1.13   & 90    & 1.89              & 2.04$\times$1.24   \\
\, 22134--VLA   & 3.6       & 4.7$\pm$0.3     & 0.82$\times$0.72   & 116   & 0.20              & 0.98$\times$0.84   \\
\, 22308--VLA   & 6.0       & 203.0$\pm$13.0  & 1.90$\times$1.07   & 86    & 0.47              & 7.66$\times$5.02   \\
\,               & 3.6       & 433.0$\pm$25.0  & 1.90$\times$1.07   & 79    & 1.14              & 7.67$\times$5.07   \\
\, 23030--VLA   & 6.0       & 945.0$\pm$44.0  & 1.62$\times$1.19   & 95    & 1.40              & 14.79$\times$7.50   \\
\,               & 3.6       & 1226.0$\pm$72.0  & 1.62$\times$1.19   & 90   & 2.52              & 12.79$\times$5.52   \\
\,               & 1.3       & 1670.0$\pm$110.0  & 1.62$\times$1.19   & 96   & 5.86              & 12.19$\times$5.65   \\
\bottomrule
\end{tabular}
\end{table}

\begin{table}[!htp]
\centering
\scriptsize
 \setlength{\tabnotewidth}{\columnwidth}
  \tablecols{8}
\caption{Physical Parameters of the sources using their 3.6~cm continuum emission}
\label{tab:tab5}
\begin{tabular}{lccccccc}
\toprule
\, VLA 3.6~cm          & Size & EM                   & n$_{\rm e}$       & M$_{\hii}$    &  N$_{\rm i}$ & Spectral & \hii region\tabnotemark{a} \\
\, Source              & (pc) & (10$^6$cm$^{-6}$ pc) & (10$^3$cm$^{-3}$) & (M$_{\odot}$) &   (s$^{-1}$) & Type     & Type                        \\
\hline
\, 01045--VLA                & 0.16 &   21.30 & 11.68 & 0.5792 & 48.39 & O8 & C \\
\, 01133--VLA                & 0.03 &   0.67 & 4.80 & 0.0015 & 45.41 & B1 & UC \\
\, 03235--VLA                & 0.03 &   2.54 & 9.72 & 0.0025 & 45.93 & B0.5 & UC \\
\, 04324--VLA                & 0.24 &   0.88 & 1.91 & 0.3501 & 47.38 & B0 & C \\
\, 04366--VLA                & 0.07 &   0.53 & 2.71 & 0.0129 & 46.10 & B0.5 & UC \\
\, 05305--VLA                & 0.06 &   0.09 & 1.18 & 0.0039 & 45.22 & B1 & UC \\
\, 05358--VLA1\tabnotemark{b} & 0.02 &  0.29 &  4.03 &  0.0003 & 44.64 & B2 & UC  \\
\, 05358--VLA2               & 0.02 &   0.14 &  2.92 &  0.0002 & 44.27 & B2 & UC \\
\, 05553--VLA                & 0.01 &   0.57 & 10.08 & 0.00002 & 43.93 & B3 & UC \\
\, 06055--VLA                & 0.01 &   0.47 & 5.61 & 0.0002 & 44.69 & B2 & UC \\
\, 06412--VLA                & 0.41 &   3.24 & 2.82 & 2.4979 & 48.40 & O8 & C \\
\, 06567--VLA                & 0.04 &   1.89 & 6.85 & 0.0058 & 46.16 & B0.5 & UC \\
\, 07299--VLA                & 0.01 &   0.13 & 4.10 & 0.00003 & 43.57 & B3 & UC \\
\, 07311--VLA                & 0.22 &   0.08 & 0.60 & 0.0873 & 46.27 & B0.5 & C \\
\, 07427--VLA                & 0.05 &   0.82 & 4.21 & 0.0054 & 45.91 & B0.5 & UC \\
\, 07528--VLA                & 0.03 &   0.58 & 4.72 & 0.0011 & 45.26 & B1 & UC \\
\, 08189--VLA                & 0.79 &   0.02 &  0.17 & 1.1159 & 46.84 & B0 & UC \\
\, 18094--G12.20             & 0.68 &   0.05 & 0.26 & 1.0963 & 47.02 & B0 & UC \\
\, 19120--G45.47             & 0.33 &   1.17 & 1.90 & 0.8519 & 47.76 & O9.5 & C \\
\, 21306--VLA                & 0.05 &   2.84 & 7.19 & 0.0155 & 46.60 & B0.5 & UC \\
\, 21334--VLA                & 0.04 &   1.66 & 6.74 & 0.0043 & 46.02 & B0.5 & UC \\
\, 21413--VLA                & 0.07 &   39.41 & 24.39 & 0.0924 & 47.91 & O9.5 & UC \\
\, 22134--VLA                & 0.01 &   3.78 & 19.30 & 0.0003 & 45.26 & B1 & UC \\
\, 22308--VLA                & 0.18 &   7.10 & 6.35 & 0.4514 & 48.01 & O9 & C \\
\, 23030--VLA                & 0.20 &   9.73 & 7.06 & 0.6858 & 48.24 & O8.5 & C \\
\bottomrule
\tabnotetext{a}{UC = \uchii region and C = Compact \hiir.} 
\tabnotetext{b}{This source has an elongated, jet-like morphology. See Appendix \ref{sec:ap-A}.}
\end{tabular}
\end{table}

\begin{table}[!htp]
\centering
\scriptsize
 \setlength{\tabnotewidth}{\columnwidth}
  \tablecols{7}
\caption{Spectral Index of the sources detected at two and/or three wavelengths}
\label{tab:spectral-index}
\begin{tabular}{lcc}
\toprule
\, VLA 3.6~cm & Wavelength & Spectral    \\
\,  Source       &   (cm)     & Index       \\
\hline
\, 01045--VLA  &   3.6 \& 6        &  1.3$\pm$0.2   \\
\, 01133--VLA  &   3.6 \& 6        &  0.6$\pm$0.8    \\
\, 03235--VLA  &   3.6 \& 6        &  2.4$\pm$0.2    \\
\, 04324--VLA  &   3.6 \& 6        &  1.2$\pm$0.2    \\
\, 04366--VLA  &   3.6 \& 6        &  1.9$\pm$0.5    \\
\, 05305--VLA  &   3.6 \& 6        &  --1.2$\pm$0.5   \\
\, 06412--VLA  &   1.3, 3.6 \& 6   &  --0.2$\pm$0.4   \\
\, 06567--VLA  &   3.6 \& 6        &  --0.5$\pm$0.3   \\
\, 07528--VLA  &   3.6 \& 6        &  0.2$\pm$0.4    \\
\, 08189--VLA  &   1.3 \& 3.6      &  0.1$\pm$0.3    \\
\, 21306--VLA  &   3.6 \& 6        &  --1.1$\pm$0.3   \\
\, 21334--VLA  &   3.6 \& 6        &  --0.6$\pm$0.1   \\
\, 21413--VLA  &   1.3, 3.6 \& 6   &  0.9$\pm$0.2    \\
\, 22308--VLA  &   3.6 \& 6        &  1.4$\pm$0.4    \\
\, 23030--VLA  &   1.3, 3.6 \& 6   &   0.4$\pm$0.3   \\
\bottomrule
\end{tabular}
\end{table}

\begin{table}[!htp]
\centering
\scriptsize
 \setlength{\tabnotewidth}{\columnwidth}
  \tablecols{7}
\caption{Physical Parameters: \hii Region Models}
\label{tab:models}
\begin{tabular}{lccccccc}
\toprule
\, VLA 3.6~cm     & Size\tabnotemark{a}  & EM                 &    n$_{\rm e}$       &  N$_{\rm i}$ & Spectral & Morphology & \hii\tabnotemark{b} \\
\, Source         & (pc) & $cm^{-6}$ pc       &  $cm^{-3}$  &  (s$^{-1}$)  & Type     &            & Type \\
\hline
\, 01045-VLA  & 0.16     &      4.39$\times$10$^{8}$  & 8.92$\times10^4$ &       6.25$\times10^{48}$    &      O6.5      & Spherical   &  UC \\
\, 01133-VLA  & 0.03    &      1.60$\times$10$^{8}$  & 2.51$\times10^5$ &       4.80$\times10^{45}$    &      B1        & Spherical  &  UC \\
\, 03235-VLA  & 0.03    &      6.68$\times$10$^{9}$  & 1.63$\times10^6$ &       1.97$\times10^{47}$    &      B0        & Spherical  &  UC \\
\, 04324-VLA  & 0.24     &      4.67$\times$10$^{8}$  & 3.87$\times10^5$ &       6.11$\times10^{47}$    &      O9.5      & Gaussian  &  C \\
\, 04366-VLA  & 0.07    &      1.37$\times$10$^{9}$  & 6.41$\times10^5$ &       7.06$\times10^{46}$    &      B0        & Spherical  &  UC \\
\, 06412-VLA  & 0.41     &      1.19$\times$10$^{6}$  & 2.88$\times10^3$ &       3.26$\times10^{48}$    &      O7.5      & Gaussian  &  {\bf  UC\tabnotemark{c} }  \\
\, 07528-VLA  & 0.03    &      3.59$\times$10$^{7}$  & 1.02$\times10^5$ &       1.96$\times10^{45}$    &      B1        & Cylindrical    & UC \\
\, 08189-VLA  & 0.79     &      1.71$\times$10$^{8}$  & 1.16$\times10^5$ &       1.30$\times10^{47}$    &      B0        & Spherical & UC \\
\, 21413-VLA  & 0.07     &      5.75$\times$10$^{8}$  & 3.10$\times10^5$ &       2.76$\times10^{48}$    &      O8        & Gaussian  & UC \\
\, 22308-VLA  & 0.18     &      9.14$\times$10$^{8}$  & 4.04$\times10^5$ &       3.85$\times10^{48}$    &      O7.5      & Gaussian  & UC \\
\, 23030-VLA  & 0.20     &      1.07$\times$10$^{8}$  & 8.85$\times10^4$ &       2.67$\times10^{48}$    &      O8        & Gaussian  & UC \\
\bottomrule
\tabnotetext{a}{Taken from the 3.6 cm RC emission} 
\tabnotetext{b}{ UC = \uchii region and C = Compact \hiir.}
\tabnotetext{c}{ UC with cometary morphology} 
\end{tabular}
\end{table}

\begin{table}[!htp]
\centering
\scriptsize
 \setlength{\tabnotewidth}{\columnwidth}
  \tablecols{7}
\caption{Final catalog of \hii regions\tabnotemark{a}} 
\label{tab:tab13}
\begin{tabular}{cccccccc}
\toprule
 IRAS & VLA 3.6~cm & Size & Size & EM & n$_{\rm e}$ & M$_{\rm \hii}$ & \hii\tabnotemark{a}  \\ 
 Source & Source & (arcsec) & (pc) & (10$^6$cm$^{-6}$ pc) & (10$^3$cm$^{-3}$) & (M$_{\odot}$) & Type   \\
\hline
01045+6505 & 01045--VLA & 3.02 & 0.16 & 21.30 & 11.68 &  0.5792 &  UC  \\
01133+6434 & 01133--VLA & 0.83 & 0.03 & 0.67 & 4.80 & 0.0015 &  UC  \\
03235+5808 & 03235--VLA & 0.87 & 0.03 & 2.54 & 9.72 & 0.0025 & UC  \\
04324+5106 & 04324--VLA & 8.56 & 0.24 & 0.88 & 1.91 & 0.3501 & C  \\
04366+5022 & 04366--VLA & 2.30 & 0.07 & 0.53 & 2.71 & 0.0129 & UC  \\
05358+3543 & 05358--VLA2 & 1.84 & 0.02 & 0.14 & 2.92 & 0.0003 & UC  \\
05553+1631 & 05553--VLA & 0.95 & 0.01 & 0.57 & 10.08 & 0.00002 & UC  \\
06055+2039 & 06055--VLA & 0.95 & 0.01 & 0.47 & 5.61 & 0.0002 & UC  \\
{\bf 06412--0105} & 06412--VLA & 0.41 & 3.24  & 1.19  & 2.82 & 2.4979 & UC\tabnotemark{b} \\
07299--1651 & 07299--VLA & 1.11 & 0.01 & 0.13 & 4.10 & 0.00003 & UC  \\
07311--2204 & 07311--VLA & 5.75 & 0.22 & 0.08 & 0.60 & 0.0873 & C  \\
07427--2400 & 07427--VLA & 1.31 & 0.05 & 0.82 & 4.21 & 0.0054 & UC  \\
07528--3441 & 07528--VLA & 3.78 & 0.03 & 0.58  & 4.72 & 0.0011 & UC \\ 
08189--3602 & 08189--VLA & 3.48 & 0.79 & 0.02  & 0.17 & 1.1159 & UC  \\
18094--1823 & 18094--G12.20 & 2.10 & 0.68 & 0.05 & 0.26 & 1.0963 & UC \\
19120--1103 & 19120--G45.47 & 2.20 & 0.33 & 1.17 & 1.90 & 0.8519 & C  \\
21413+5442 & 21413--VLA & 1.70 & 0.07 & 39.41 & 24.39 & 0.0924 & UC  \\
22134+5834 & 22134--VLA & 0.91 & 0.01 & 3.78 & 19.30 & 0.0003 & UC  \\
22308+5812 & 22308--VLA & 6.24 & 0.18 & 7.10 & 6.35 & 0.4514 & UC  \\
23030+5958 & 23030--VLA & 4.30 & 0.20 & 9.73  & 7.06 & 0.6858 & UC \\
\bottomrule
\tabnotetext{a}{ UC = \uchii region and C = Compact \hiir.} 
\tabnotetext{b}{ UC with cometary morphology} 
\end{tabular}
\end{table}

\clearpage


\end{document}